\documentclass[%
reprint,
superscriptaddress,
amsmath,amssymb,
pre
]{revtex4-1}

\usepackage{graphicx}
\usepackage{dcolumn}
\usepackage{bm}
\usepackage{bbm}
\usepackage[colorlinks=true,allcolors=blue]{hyperref}
\usepackage[mathscr]{eucal}
\usepackage{tikz}
\usetikzlibrary{shapes,arrows,backgrounds,patterns,decorations.markings}
\usepackage{longtable}
\usepackage{booktabs}
\usepackage{hhline}
\usepackage{mathtools}
\usepackage{balance}
\usepackage{float}
\usepackage{makecell}

\definecolor{blue}{RGB}{61, 125, 227}

\allowdisplaybreaks 

\begin{document}

\title{Weighted hypersoft configuration model
}
\author{Ivan Voitalov}
\affiliation{Department of Physics, Northeastern University, Boston, Massachusetts 02115, USA}
\affiliation{Network Science Institute, Northeastern University, Boston, Massachusetts 02115, USA}
\author{Pim van der Hoorn}
\affiliation{Department of Mathematics and Computer Science, Eindhoven University of Technology, Postbus 513, 5600 MB Eindhoven, Netherlands}
\author{Maksim Kitsak}
\affiliation{Department of Physics, Northeastern University, Boston, Massachusetts 02115, USA}
\affiliation{Network Science Institute, Northeastern University, Boston, Massachusetts 02115, USA}
\affiliation{Faculty of Electrical Engineering, Mathematics and Computer Science, Delft University of Technology, 2628 CD, Delft, Netherlands}
\author{Fragkiskos Papadopoulos}
\affiliation{Department of Electrical Engineering, Computer Engineering and Informatics, Cyprus University of Technology, 33 Saripolou Street, 3036 Limassol, Cyprus}
\author{Dmitri Krioukov}
\affiliation{Department of Physics, Northeastern University, Boston, Massachusetts 02115, USA}
\affiliation{Network Science Institute, Northeastern University, Boston, Massachusetts 02115, USA}
\affiliation{Department of Mathematics, Northeastern University, Boston, Massachusetts 02115, USA}
\affiliation{\mbox{Department of Electrical and Computer Engineering, Northeastern University, Boston, Massachusetts 02115, USA}}

\pacs{}

\begin{abstract}

Maximum entropy null models of networks come in different flavors that depend on the type of constraints under which entropy is maximized. If the constraints are on degree sequences or distributions, we are dealing with configuration models. If the degree sequence is constrained exactly, the corresponding microcanonical ensemble of random graphs with a given degree sequence is the configuration model {\it per se}. If the degree sequence is constrained only on average, the corresponding grand-canonical ensemble of random graphs with a given expected degree sequence is the soft configuration model. If the degree sequence is not fixed at all but randomly drawn from a fixed distribution, the corresponding hypercanonical ensemble of random graphs with a given degree distribution is the hypersoft configuration model, a more adequate description of dynamic real-world networks in which degree sequences are never fixed but degree distributions often stay stable. Here, we introduce the hypersoft configuration model of weighted networks. The main contribution is a particular version of the model with power-law degree and strength distributions, and superlinear scaling of strengths with degrees, mimicking the properties of some real-world networks. As a byproduct, we generalize the notions of sparse graphons and their entropy to weighted networks.

\end{abstract}

\maketitle

\section{Introduction}\label{sec:intro}

Many real-world complex systems that can be represented as networks~\cite{newman2018networks,barabasi2016network} require weighted representations in which connections between nodes are characterized by positive weights~\cite{barrat2004architecture}. For example, in modeling the global spread of an epidemic using an air transportation network as a backbone, it is important to know not only that there exists a flight from airport~$i$ to airport~$j$, but also the volume of the passenger flow between the two airports. This volume is usually encoded as the link weight $w_{ij}$~\cite{colizza2006role,brockmann2013hidden}.

Within the plethora of weighted and unweighted network models developed in network science to study the structure and function of real-world networks, \textit{maximum-entropy models}~\cite{park2004statistical,bianconi2007entropy,garlaschelli2008maximum,garlaschelli2009generalized,squartini2011analytical,sagarra2013statistical,mastrandrea2014enhanced,squartini2015unbiased,van2018sparse,gabrielli2019grand,cimini2019statistical} play a special role. They serve as null models that are indispensable in studying the intricate interdependencies between different network properties~\cite{vazquez2004topological,foster2011clustering,colomer2013deciphering,orsini2015quantifying,maslov2002specificity,park2003origin,guimera2007classes,opsahl2008prominence,colizza2006detecting,amaral2006lies,zhou2004rich}. Within this class of models, perhaps the most studied are classical random graphs~\cite{solomonoff1951connectivity,gilbert1959random,erdos1959random,erdos1960evolution} that maximize ensemble entropy with the average degree constrained to a given value. They do not reproduce heterogeneous degree distributions observed in many real-world networks~\cite{voitalov2019scale}, which motivated the development of the configuration model.

The \textit{configuration model (CM)}~\cite{bender1978asymptotic,molloy1995critical} is a microcanonical ensemble of random graphs with \textit{sharp constraints} on the degree sequence, meaning that every graph in this ensemble has exactly the same degree sequence, e.g., the one observed in a real-world network. In the CM, every graph satisfying the constraint has the same probability in the ensemble. All other graphs are excluded.
The \textit{soft configuration model (SCM)}~\cite{chung2002connected,chung2002average,park2004statistical,garlaschelli2008maximum,squartini2011analytical} is a grand-canonical ensemble of random graphs with \textit{soft constraints} on the degree sequence. This means that the \textit{expected} degree sequence in the ensemble is equal to a given degree sequence.

In both CM and SCM, a fixed degree sequence is the constraint under which the ensemble entropy is maximized, either micro- or grand-canonically. This constraint, however, does not properly reflect the dynamic nature of node degrees observed in many real networks, where the degrees of all nodes may constantly change, while the shape of the degree distribution stays stable~\cite{newman2001clustering,dhamdhere2011twelve}. These observations motivated the development of the hypersoft configuration model.

The \textit{hypersoft configuration model (HSCM)}~\cite{caldarelli2002scale,boguna2003class,anand2014entropy,van2018sparse} is a hypercanonical ensemble of random graphs whose entropy is maximized under the constraint that the \textit{degree distribution} has a given shape. The HSCM belongs to the class of models with hidden variables~\cite{boguna2003class}, meaning that each node has a latent parameter sampled from a fixed distribution. This latent distribution defines the degree distribution, so that by tuning the former, one can reproduce any shape of the latter.

The ((H)S)CM story outlined above for unweighted networks finds an incomplete and somewhat distorted reflection in weighted networks where the constraints under which entropy is maximized are on both degrees and strengths of nodes~\cite{mastrandrea2014enhanced}. The \textit{weighted configuration model (WCM)} is a microcanonical ensemble of networks with sharp constraints on both the degree and strength sequences. Every weighted graph in this ensemble has the same degree and strength sequences, and the same probability in the ensemble. However, we note that several models that come under the WCM name~\cite{serrano2005weighted,britton2011weighted,mastrandrea2014enhanced} are not really as defined above. The \textit{weighted soft configuration model (WSCM)}~\cite{garlaschelli2009generalized,mastrandrea2014enhanced,squartini2015unbiased,gabrielli2019grand} is a grand-canonical ensemble of networks with soft constraints on the degree and strength sequences, meaning that the expected degree and strength sequences in the ensemble are equal to given degree and strength sequences.

In this paper, we introduce the \textit{weighted hypersoft configuration model (WHSCM)} which is a hypercanonical ensemble of networks with a fixed \textit{joint distribution} of degrees and strengths. Similar to the HSCM, the WHSCM is a hidden variable model where each node has two latent parameters sampled from a fixed joint distribution which defines the joint distribution of degrees and strengths. We summarize the taxonomy of the (((W)H)S)CM models in Table~\ref{tab:models}.

\begin{longtable}[h]{p{0.25\textwidth} p{0.11\textwidth} p{0.11\textwidth}}
    \caption{Configuration models of unweighted and weighted networks. The subject of this paper is the WHSCM.
    }\label{tab:models}\\
    \hhline{===}
    \textbf{Maximum entropy\newline constraints}   &     \textbf{Unweighted models}     &     \textbf{Weighted models} \\
    \hline
    \endhead

    \rule{0pt}{3ex}Exact degree (and strength)\newline sequence    &     CM                            &     WCM \\
    Expected degree (and strength)\newline sequence &     SCM                           &     WSCM \\
    Degree (and strength)\newline distribution      &     HSCM                          &     \textbf{WHSCM} \\
    \hhline{===}
\end{longtable}

Besides introducing the WHSCM in general, the main focus of this paper is a much more involved task, which is to identify the joint distribution of latent parameters
that reproduces several features of degree and strength distributions observed in many real weighted networks~\cite{barrat2004architecture,garlaschelli2005scale,bagler2008analysis,barthelemy2011spatial,popovic2012geometric,allard2017geometric}. Specifically, these features are: (1)~\textit{power-law degree distribution}, (2)~\textit{super-linear scaling between strengths and degrees}, and (3)~\textit{sparsity}. The last one means that the average degree is constant as a function of the network size.

We proceed by first providing all the necessary motivation and background information in Sec.~\ref{sec:background} that ends with the introduction of the most general form of the WHSCM. In Sec.~\ref{sec:requirements}, we document in detail the real-world-network-dictated properties, mentioned above, that we want our particular power-law version of the WHSCM to reproduce. To reproduce those properties, we need some experimental input from the WSCM as defined in Ref.~\cite{gabrielli2019grand}, a subject of Sec.~\ref{sec:mle}. Based on this input, we derive in Sec.~\ref{sec:whscm} the WHSCM latent parameter distribution that satisfies the requirements of Sec.~\ref{sec:requirements}. We check in simulations that these requirements are indeed satisfied in Sec.~\ref{sec:synthetic-nets}. To demonstrate how the constructed model can be used in the analysis of real-world networks, we juxtapose several real-world networks and their WHSCM counterparts in Sec.~\ref{sec:real-nets}. We conclude in Sec.~\ref{sec:discussion} with the discussion of obvious and less obvious limitations, caveats, wishful thoughts, and abstract remarks. We release our implementation of the WHSCM graph generator in a software package available at GitHub~\cite{githubcode}.

\section{Motivation, background, and general WHSCM}\label{sec:background}

\subsection{Maximum entropy models}\label{ssec:max-ent-models-background}

Understanding mechanisms that drive formation of complex networks and dynamical processes running in them is a crucial task in network science~\cite{barrat2008dynamical}. It is commonly believed that many real complex systems are \textit{self-organizing}, i.e., adjusting their structure to optimize their function~\cite{newman2003structure}. Because of that, a lot of past research was dedicated to finding structural network properties that may be indicative of yet unknown optimization mechanisms behind network evolution and function~\cite{newman2003structure,motter2007introduction,burda2011motifs}. However, due to potentially strong interdependencies between different structural properties of networks~\cite{vazquez2004topological,foster2011clustering,colomer2013deciphering,orsini2015quantifying}, it is important to ensure that a property of interest is indeed a salient feature, and not a mere consequence of some of its other structural properties. A method that is often used to make this check is to compare the significance of the structural property present in a given network with respect to the same property in benchmark null model networks~\cite{maslov2002specificity,park2003origin,guimera2007classes,opsahl2008prominence}. This step should be taken with care as choosing an inappropriate null model for a given network may lead to wrong conclusions about its functional and structural features~\cite{colizza2006detecting,amaral2006lies,zhou2004rich}.

The maximum entropy network null models~\cite{park2004statistical,garlaschelli2008maximum,garlaschelli2009generalized,anand2009entropy,mastrandrea2014enhanced,squartini2015unbiased,cimini2019statistical} have proven to be an indispensable tool in avoiding possible statistical biases caused by interdependencies of structural network properties. These models are ensembles of networks that reproduce given structural properties, and that are maximally random in all other respects. This maximal randomness is important for a great variety of tasks~\cite{vazquez2004topological,foster2011clustering,colomer2013deciphering,orsini2015quantifying,maslov2002specificity,park2003origin,guimera2007classes,opsahl2008prominence,colizza2006detecting,amaral2006lies,zhou2004rich}. For a basic example, if a maximum entropy null model is defined by property~$X$ observed in a real-world network, and if the model also reproduces some other property~$Y$ of the network, then we know right away that $Y$ is not a salient independent feature, but a statistical consequence of~$X$~\cite{orsini2015quantifying}.

Maximum entropy network models are usually formulated for a given observed network $G^{\ast}$ of size $n$ in terms of sets of \textit{constraints}. These constraints are usually some properties $C(G^{\ast})$ of network $G^{\ast}$ that the model is required to reproduce. We note that the constraints do certainly \emph{not} have to be properties of any real network, they can be any set of (artificial) network properties, but what we describe is the most common application scenario.

Given constraints $C(G^{\ast})$, the model is then an ensemble of graphs $\mathcal{G}$ of the same size $n$ as the original graph $G^{\ast}$, with each graph $G \in \mathcal{G}$ appearing in the ensemble with probability $P(G)$, known as the ensemble distribution. The ensemble distribution in a maximum entropy model defined by constraints $C(G^{\ast})$ is the unique unbiased probability distribution $P(G)$ that maximizes Gibbs/Shannon entropy
\begin{equation}\label{eq:shannon-entropy}
    S = -\sum\limits_{G \in \mathcal{G}} P(G) \log{P(G)},
\end{equation}
and that satisfies the constraints $C(G^{\ast})$ and the normalization condition $\sum_{G \in \mathcal{G}} P(G) = 1$~\cite{jaynes1957information,cimini2019statistical}.

\subsection{Unweighted configuration models}\label{ssec:max-ent-unweighted-background}

In the simplest case of \textit{undirected} and \textit{unweighted} networks, the degrees of all nodes in a given network are frequently used as constraints in maximum entropy null models. The simplest example is the configuration model.

\textbf{Configuration model (CM)~\cite{bender1978asymptotic,molloy1995critical}.} The CM is a microcanonical ensemble of graphs with the same degree sequence as observed in a real network. That is, given the degree sequence $\{k_1^{\ast},\ldots,k_{n}^{\ast}\} = \boldsymbol{k^{\ast}}$ observed in a graph $G^{\ast}$, the CM ensemble consists of all graphs $G$ with exactly the same degree sequence, i.e., for each degree sequence $\boldsymbol{k} (G)$ of an ensemble graph $G$, the following holds:
\begin{equation}\label{eq:cm-constraint}
    \boldsymbol{k}(G) = \boldsymbol{k^{\ast}}.
\end{equation}
The distribution $P(G)$ that maximizes Shannon entropy in Eq.~\eqref{eq:shannon-entropy} is the uniform distribution over the set of all graphs whose degree sequence is $\boldsymbol{k^{\ast}}$, meaning that for these graphs $P(G)=1/\mathcal{N}_{\boldsymbol{k^{\ast}}}$ and $S=\log\mathcal{N}_{\boldsymbol{k^{\ast}}}$, where $\mathcal{N}_{\boldsymbol{k^{\ast}}}$ is the number of graphs with the degree sequence $\boldsymbol{k^{\ast}}$. For all other graphs, $P(G)=0$.

\textbf{Soft configuration model (SCM)~\cite{chung2002connected,chung2002average,park2004statistical,garlaschelli2008maximum,squartini2011analytical}.} The SCM is a grand-canonical ensemble of graphs whose expected degree sequence is constrained to a given (observed) sequence. Specifically, given a degree sequence $\boldsymbol{k^{\ast}}$, the SCM constraint is
\begin{equation}
    \sum\limits_{G \in \mathcal{G}} \boldsymbol{k}(G) P(G) = \boldsymbol{k^{\ast}},
\end{equation}
where $\boldsymbol{k}(G)$ is the degree sequence in an ensemble graph $G$. Since the degree sequence used as a constraint is reproduced only in expectation, it does not have to consist of integers only; the expected degrees can be any positive real numbers.

As in any (grand)canonical ensemble, the Shannon entropy in the SCM is usually maximized using the method of Lagrange multipliers~\cite{park2004statistical}, yielding the familiar Gibbs/Boltzmann exponential family distribution
\begin{equation}
    P(G) = \frac{e^{-H_{SCM}(G)}}{Z}
\end{equation}
with Hamiltonian
\begin{equation}\label{eq:scm-hamiltonian}
    H_{SCM}(G) = \sum\limits_{(i,j) \in \mathcal{P}} (\nu_i + \nu_j) a_{ij}=\sum_i\nu_ik_i(G),
\end{equation}
where $\nu_i$ is the Lagrange multiplier coupled to node~$i$, $k_i(G)$ $i$'s degree in $G$, $\mathcal{P}$ the set of all node pairs, $a_{ij}$ the adjacency matrix of~$G$, and $Z = \sum_{G \in \mathcal{G}} e^{-H_{SCM}(G)}$ the partition function. In statistical terms, these equations say that the degree sequence is the sufficient statistics in the ensemble, so that all graphs with the same degree sequence have the same probability in the ensemble.

Graphs~$G$ can be sampled from the ensemble distribution $P(G)$ constructively by walking over all node pairs $i,j$ and linking them with the Fermi-Dirac \textit{connection probability}
\begin{equation}\label{eq:scm-conn-prob}
    p_{ij}=p(\nu_i, \nu_j) = \frac{1}{1 + e^{\nu_i + \nu_j}}.
\end{equation}
The expected degree $\kappa_i$ of node $i$ in the ensemble is thus
\begin{equation}\label{eq:scm-kappa-i}
    \kappa_i=\sum_jp_{ij},
\end{equation}
so that the values of the Lagrange multipliers $\nu_i$ for a given $\boldsymbol{k^{\ast}}$ are found as the solution of the system of $n$ equations
\begin{equation}\label{eq:scm-sol-nu}
    \kappa_i=k_i^\ast.
\end{equation}

\textbf{Hypersoft configuration model (HSCM)~\cite{caldarelli2002scale,boguna2003class,anand2014entropy,van2018sparse}.} The HSCM is a hypercanonical ensemble of graphs defined by the constraint that the expected degree distribution has a given form. The HSCM can be viewed as a hyperparametrization of the SCM, in that the Lagrange multipliers $\nu$ are not fixed by any degree sequence as solutions of \eqref{eq:scm-sol-nu}, but random, sampled independently for each node $i$ from a fixed distribution $\rho(\nu)$:
\begin{equation}
    \nu_i \leftarrow \rho(\nu).
\end{equation}
Having a sampled sequence of Lagrange multipliers $\nu_i$, the nodes are then connected as in the SCM, with the Fermi-Dirac connection probability in Eq.~\eqref{eq:scm-conn-prob}.

The expected degree $\kappa(\nu)$ of a node with Lagrange multiplier~$\nu$ in the ensemble is
\begin{equation}\label{eq:hscm-kappa(nu)}
  \kappa(\nu) = n\int p(\nu,\nu') \rho(\nu')\,d\nu',
\end{equation}
where $p(\nu,\nu')$ is from Eq.~\eqref{eq:scm-conn-prob}.
It is convenient to abuse the notations by defining the expected degree random variable $\kappa$ via
\begin{equation}\label{eq:hscm-kappa-abuse}
  \kappa=\kappa(\nu),
\end{equation}
where the left-hand side (l.h.s.) $\kappa$ is a random variable, but the right-hand side (r.h.s.) $\kappa(\nu)$ is a function, defined in Eq.~\eqref{eq:hscm-kappa(nu)}, of the random variable $\nu$ whose distribution is $\rho(\nu)$. That is, the last equation is a change of latent variables from $\nu$ to $\kappa$.
One can show~\cite{van2018sparse} that in sparse graphs the $\kappa(\nu)$ function can be well approximated as
\begin{equation}\label{eq:hscm-kappa(nu)-approx}
  \kappa(\nu)=\kappa_0e^{R-\nu}=\sqrt{\bar{k}n}\,e^{-\nu},
\end{equation}
where $R$ and $\kappa_0$ are the interchangeable parameters that control the expected average degree $\bar{k}=\kappa_0^2e^{2R}/n$ in the ensemble. With this approximation, the Fermi-Dirac connection probability in Eq.~\eqref{eq:scm-conn-prob} can be rewritten in terms of the $\kappa$ variables as
\begin{equation}\label{eq:scm-conn-prob-kappa}
  p(\kappa_i,\kappa_j)=\frac{1}{1+\frac{\bar{k}n}{\kappa_i\kappa_j}}.
\end{equation}
As proven in~\cite{janson2010asymptotic}, the classical limit (or Chung-Lu~\cite{chung2002average,chung2002connected}) approximation 
\begin{equation}\label{eq:scm-conn-prob-cl}
  p_{cl}(\kappa_i,\kappa_j)=\min\left(1,\frac{\kappa_i\kappa_j}{\bar{k}n}\right)
\end{equation}
to the connection probability~\eqref{eq:scm-conn-prob-kappa} ``almost always works,'' in the sense that the HSCM ensembles of random graphs defined by the connections probabilities~\eqref{eq:scm-conn-prob-kappa} and~\eqref{eq:scm-conn-prob-cl} are asymptotically equivalent under very mild assumptions on the distribution of $\kappa$.

Since Eq.~\eqref{eq:hscm-kappa-abuse} defines the relation between the two random variables $\kappa$ and $\nu$, it also defines the relation between their distributions $\rho(\nu)$ and $\rho(\kappa)$ via the standard formula
\begin{equation}
  \rho_\kappa(\kappa)=\rho_\nu[\nu(\kappa)]\left|\nu'(\kappa)\right|,
\end{equation}
where $\nu(\kappa)$ is the inverse function of $\kappa(\nu)$ and $\nu'(\kappa)$ its derivative.
By the definition of $\kappa$, its distribution $\rho(\kappa)$ is the distribution of expected degrees in the HSCM, the analogy of the expected degree sequence $\kappa_i$~\eqref{eq:scm-kappa-i} in the SCM. Therefore, the HSCM analogy of the SCM constraints~\eqref{eq:scm-sol-nu} is
\begin{equation}\label{eq:hscm-constraints}
  \rho(\kappa)=\rho^\ast(\kappa),
\end{equation}
where $\rho^\ast(\kappa)$ is any desired expected degree distribution. For example, it can be a pure power law, i.e., the continuous Pareto distribution
\begin{equation}
  \rho(\kappa)=(\gamma-1)\kappa_0^{\gamma-1}\kappa^{-\gamma},\quad\kappa>\kappa_0>0.
\end{equation}
In view of Eq.~\eqref{eq:hscm-kappa(nu)-approx}, the distribution of the Lagrange multipliers $\nu$ is exponential in this case:
\begin{equation}\label{eq:hscm-rho}
    \rho(\nu) = (\gamma - 1) e^{(\gamma - 1) (\nu - R)}, \quad \nu\in(-\infty,R].
\end{equation}

The general exact expression for the expected average degree in the ensemble is
\begin{equation}\label{eq:hscm-average-degree}
  \bar{k}=\bar{\kappa}=\int\kappa\rho(\kappa)\,d\kappa=\int\kappa(\nu)\rho(\nu)\,d\nu.
\end{equation}

In sparse graphs, the expected degree distribution $\rho(\kappa)$ defines the degree distribution $P(k)$ via
\begin{equation}\label{eq:hscm-P(k)}
  P(k) = \int \mathrm{Pois}(k|\kappa)\rho(\kappa)\,d\kappa,
\end{equation}
where $\mathrm{Pois}(k|\kappa)$ is the Poisson distribution with mean~$\kappa$~\cite{boguna2003class,britton2006generating,bollobas2007phase}. The Poisson distribution appears here as the $n\to\infty$ limit of the distributions of sums of $n$ Bernoullis with random rates~$p_{ij}$ whose sum $\sum_jp_{ij}$ converges to~$\kappa_i$. The distributions~$P(k)$ in the form~\eqref{eq:hscm-P(k)} are called mixed Poisson distributions with $\kappa$ the mixing parameter~\cite{grandell1997mixed}. A short proof that the degree distribution in the HSCM is a mixed Poisson distribution can be found in Th.~3.1 in Ref.~\cite{britton2006generating}, for instance. The proof relies on the observation that the generating function of the degree distribution is the generating function of the mixed Poisson distribution. The shape of a mixed Poisson distribution $P(k)$ follows the shape of the distribution of its mixing parameter $\rho(\kappa)$. For example, if $\rho(\kappa)$ is Pareto with exponent $\gamma$, then the Pareto-mixed Poisson distribution is also a power law, albeit impure, with the same exponent, $P(k)\sim k^{-\gamma}$, or in stricter terms, it is a regularly varying distribution with exponent $\gamma$~\cite{voitalov2019scale}.

As shown in Ref.~\cite{van2018sparse} and discussed in Appendix~\ref{sec:proof}, the entropy of HSCM graphs is maximized across all graphs whose degree distribution converges to a given distribution.

\begin{figure}
 \centering
 \includegraphics[width=.45\textwidth]{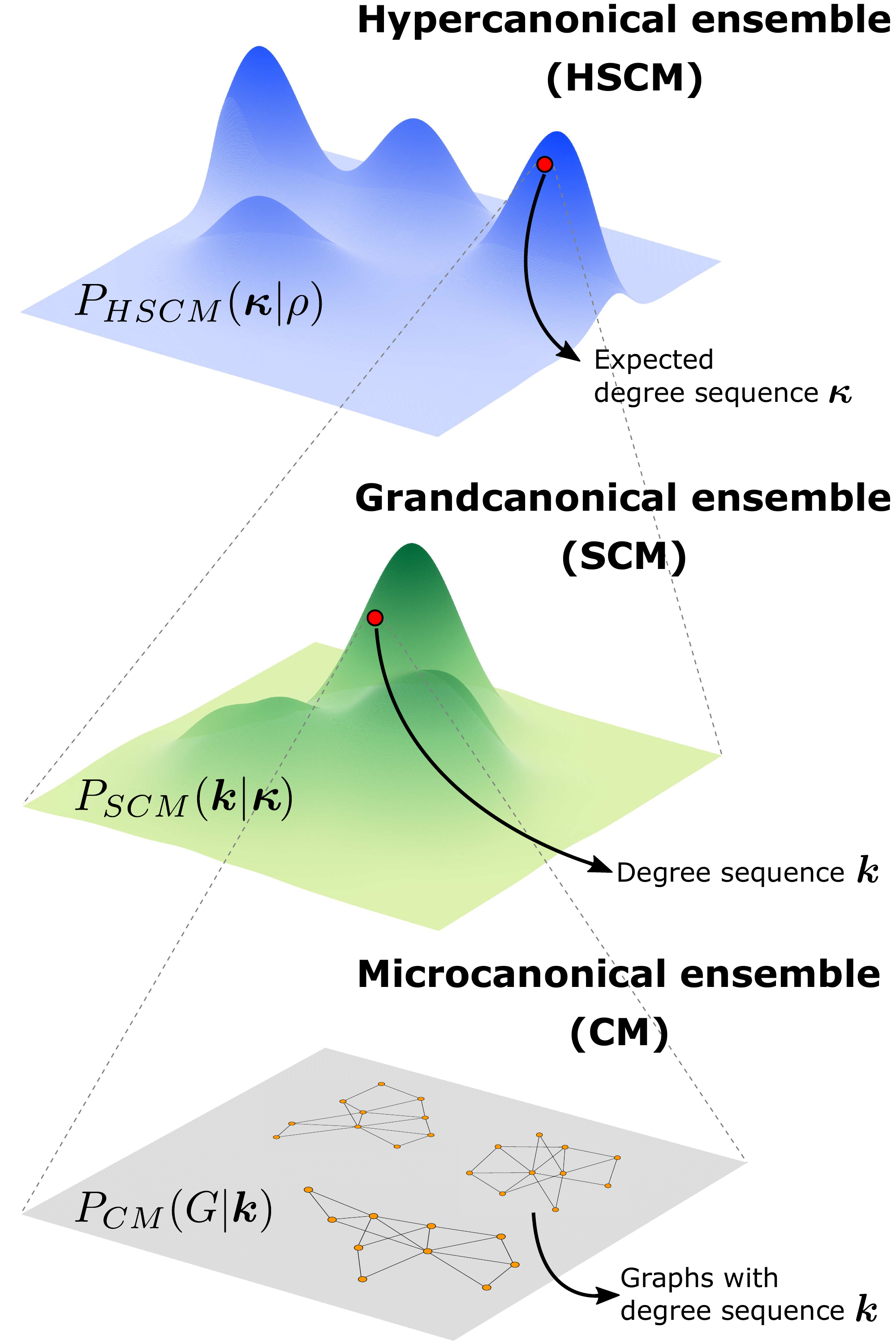}
  \caption{
  {\bf The nested hierarchy of the configuration models.}
  Sampling a graph $G$ from the hypercanonical HSCM ensemble can be done in three steps: (1)~sample a sequence $\boldsymbol{\kappa}=\{\kappa_i\}$ of expected degrees independently from the distribution $\rho(\kappa)$, i.e., from $P_{HSCM}(\boldsymbol{\kappa}|\rho)=\prod_i\rho(\kappa_i)$; (2)~sample a degree sequence $\boldsymbol{k}=\{k_i\}$ from the SCM degree sequence distribution $P_{SCM}(\boldsymbol{k}|\boldsymbol{\kappa})$; and (3)~sample graph $G$ from the uniform CM distribution $P_{CM}(G|\boldsymbol{k})$. The HSCM probability distribution can thus be written as a chain $P_{HSCM}(G|\rho)=\int_{\boldsymbol{\kappa}}\sum_{\boldsymbol{k}}P_{CM}(G|\boldsymbol{k})P_{SCM}(\boldsymbol{k}|\boldsymbol{\kappa})P_{HSCM}(\boldsymbol{\kappa}|\rho)\,d\boldsymbol{\kappa}$  
  showing that the ensemble at the next level of the hierarchy is a probabilistic mixture of the ensembles at the previous level. In the graphon theory~\cite{janson2013graphons}, one has to consider the fourth highest level (not shown) where $\rho$ is also random. Moving up in the hierarchy adds new sources of randomness, thus increasing entropy. Since $P_{SCM}(\boldsymbol{k}|\boldsymbol{\kappa})$ is intractable---it is a mixture of mixed Poisson distributions---in practice it is much easier to sample graphs as described in the text. The same picture applies to the weighted case upon the addition of strengths $\boldsymbol{s}$ and their expectations $\boldsymbol{\sigma}$.
  \label{fig:ensemble-hierarchy}
  }
\end{figure}

The nested hierarchy of the described configuration models is visualized in Fig.~\ref{fig:ensemble-hierarchy}.

\subsection{Weighted configuration models}\label{ssec:max-ent-weighted-background}

For \textit{weighted} networks, the configuration models are analogous to those for unweighted networks discussed above. They are formulated in terms of constraints on both degrees and strengths of nodes. 

\textbf{Weighted Configuration Model (WCM).} The WCM is a microcanonical ensemble of weighted networks with sharp constraints on the degree and strength sequences $\boldsymbol{k^{\ast}}$ and $\boldsymbol{s^{\ast}}$. The ensemble consists of weighted graphs that have exactly the same degree and strength sequences as in the observed graph:
\begin{align}
    \boldsymbol{k} (G) = \boldsymbol{k^{\ast}},\label{eq:wcm-k-constraint}\\
    \boldsymbol{s} (G) = \boldsymbol{s^{\ast}}.\label{eq:wcm-s-constraint}
\end{align}
Analogously to its unweighted version, the distribution maximizing Shannon entropy is the uniform distribution over all graphs with the joint degree-strength sequence equal to $(\boldsymbol{k^{\ast}},\boldsymbol{s^{\ast}})$. This ensemble is well defined and such a uniform distribution always exists because the space of weight matrices $\{w_{ij}\}$ representing graphs satisfying the constraints~(\ref{eq:wcm-k-constraint},\ref{eq:wcm-s-constraint}) is always of a finite volume (Lebesgue measure) if weights are real, or of a finite cardinality if they are integer. Indeed, since all weights are positive, they all are bounded by the minimum strength of the two incident nodes: $0<w_{ij}\leq\min(s_i^\ast,s_j^\ast)$.

We note that there exist several network models introduced under the WCM name in the past that are different from the WCM definition above. Specifically, in Ref.~\cite{serrano2005weighted} the authors consider the unweighted \textit{multigraph} CM with degree sequences following power-law distributions with $\gamma < 2$. In these settings, multiple links between the same pairs of nodes are present with high probability. These multiple links are treated as weights in Ref.~\cite{serrano2005weighted}. In Ref.~\cite{mastrandrea2014enhanced}, the WCM is a model in which only the strength sequence is fixed, but the degree sequence is not fixed. In Ref.~\cite{britton2011weighted}, the WCM is a model in which the degree sequence is fixed, while the weights of links attached to a node are sampled from a distribution that is allowed to depend on the node degree. To the best of our knowledge, the WCM as we defined it above has been introduced only in Ref.~\cite{bianconi2009entropy}, albeit under a different name.

\textbf{Weighted Soft Configuration Model (WSCM)~\cite{garlaschelli2009generalized,mastrandrea2014enhanced,squartini2015unbiased,gabrielli2019grand}.} The WSCM is a grand-canonical ensemble of networks whose expected degree and strength sequences are constrained to given (observed) sequences. For any given (observed) degree and strength sequences $\boldsymbol{k^{\ast}}$ and $\boldsymbol{s^{\ast}}$, the WSCM constraints are
\begin{align}
    \sum\limits_{G \in \mathcal{G}} \boldsymbol{k}(G) P(G) = \boldsymbol{k^{\ast}},\label{eq:wscm-k-constraint}\\ 
    \sum\limits_{G \in \mathcal{G}} \boldsymbol{s}(G) P(G) = \boldsymbol{s^{\ast}},\label{eq:wscm-s-constraint}
\end{align}
where $\boldsymbol{k}(G)$ and $\boldsymbol{s}(G)$ are the degree and strength sequences of an ensemble graph $G$. As in the unweighted case, Shannon entropy is maximized using the method of Lagrange multipliers leading to the ensemble distribution
\begin{equation}\label{eq:wscm-gibbs}
    P(G) = \frac{e^{-H_{WSCM}(G)}}{Z}
\end{equation}
with Hamiltonian
\begin{align}
    H_{WSCM}(G) &= \sum\limits_{(i,j) \in \mathcal{P}} (\nu_i + \nu_j) a_{ij} + \sum\limits_{(i,j) \in \mathcal{E}} (\mu_i + \mu_j) w_{ij}\nonumber\\
    &= \sum_i \nu_ik_i(G) + \mu_i s_i(G),\label{eq:wscm-hamiltonian}
\end{align}
where $\nu_i$ and $\mu_i$ are the Lagrange multipliers coupled to node~$i$, constraining its degree $k_i(G)$ and strength $s_i(G)$, $a_{ij}$ and $w_{ij}$ are the adjacency and weight matrices of $G$, $\mathcal{P}$ and $\mathcal{E}$ are the sets of node pairs and connected node pairs, and $Z = \sum_{G \in \mathcal{G}} e^{-H_{WSCM}(G)}$ the partition function. The sufficient statistics are thus the degree and strength sequences, so that all graphs with the same degree and strength sequences have the same probability in the ensemble.

The WSCM was defined and studied both for positive integer-~\cite{garlaschelli2009generalized,mastrandrea2014enhanced,squartini2015unbiased} and real-valued weights~\cite{gabrielli2019grand}. In the latter case the graphs can be sampled constructively from the ensemble distribution $P(G)$ as follows. First, every pair of nodes $i$ and $j$ is connected with the connection probability
\begin{equation}\label{eq:wscm-conn-prob}
    p_{ij}=p(\nu_i, \mu_i, \nu_j, \mu_j) = \frac{1}{1 + e^{\nu_i + \nu_j} (\mu_i + \mu_j)}.
\end{equation}
Second, every established link $i,j$ is weighted by a random weight $w_{ij}$ sampled from the exponential distribution with rate $\mu_i + \mu_j$:
\begin{equation}\label{eq:wscm-weight-sampling}
    w_{ij} \leftarrow \mathrm{Exp}(w|\mu_i+\mu_j)=(\mu_i+\mu_j)e^{-(\mu_i+\mu_j)w}.
\end{equation}
The expected weight of link $i,j$ is then
\begin{equation}\label{eq:whscm-omega}
  \omega_{ij} = \omega(\nu_i,\mu_i,\nu_j,\mu_j) = \frac{p(\nu_i,\mu_i,\nu_j,\mu_j)}{\mu_i+\mu_j}=\frac{p_{ij}}{\mu_i+\mu_j}.
\end{equation}
The expected degree $\kappa_i$ and strength $\sigma_i$ of node $i$ in the ensemble are thus
\begin{align}
    \kappa_i &= \sum_jp_{ij}, \label{eq:wscm-kappa-i}\\
    \sigma_i &= \sum_j\omega_{ij}, \label{eq:wscm-sigma-i}
\end{align}
so that
the Lagrange multipliers $\nu_i,\mu_i$ are found for given $\boldsymbol{k^{\ast}},\boldsymbol{s^{\ast}}$ as the solution of the system of the $2n$ equations
\begin{align}
  \kappa_i    &= k_i^\ast, \label{eq:wscm-kappa-constraints}\\
  \sigma_i    &= s_i^\ast. \label{eq:wscm-sigma-constraints}
\end{align}

\textbf{Weighted Hypersoft Configuration Model (WHSCM).} We introduce the WHSCM here as a hypercanonical ensemble of weighted networks with positive real-valued weights. The maximum entropy constraint of the model is that the joint distribution of expected degrees and strengths has a given form. Similar to the HSCM, which is a hyperparametrization of the SCM, the WHSCM is a hyperparametrization of the WSCM, meaning that the Lagrange multipliers $\nu_i,\mu_i$ of node $i$ are not fixed by any fixed degree and strength sequences. Instead, $\nu_i,\mu_i$ are random, sampled independently for each node $i$ from a fixed joint probability distribution $\rho(\nu,\mu)$:
\begin{equation}\label{eq:whscm-parameters-sampling}
    (\nu_i,\mu_i) \leftarrow \rho(\nu, \mu).
\end{equation}
Having a joint sampled sequence of Lagrange multipliers $\nu_i,\mu_i$, the nodes are then connected as in the WSCM, with the connection probability in Eq.~\eqref{eq:wscm-conn-prob}, and the established links are weighted by random weights in Eq.~\eqref{eq:wscm-weight-sampling}.

The expected degree $\kappa(\nu,\mu)$ and strength $\sigma(\nu,\mu)$ of a node with Lagrange multipliers $\nu$ and $\mu$ in the ensemble are
\begin{align}
    \kappa(\nu, \mu) &= n \iint p(\nu, \mu, \nu^{\prime}, \mu^{\prime}) \rho(\nu^{\prime}, \mu^{\prime}) \, d\nu^{\prime} \, d\mu^{\prime},\label{eq:whscm-bar(k)}\\
    \sigma(\nu, \mu) &= n \iint \omega(\nu, \mu, \nu^{\prime}, \mu^{\prime}) \rho(\nu^{\prime}, \mu^{\prime}) \, d\nu^{\prime} \, d\mu^{\prime},\label{eq:whscm-bar(s)}
\end{align}
where $p,\omega$ are from Eqs.~(\ref{eq:wscm-conn-prob},\ref{eq:whscm-omega}).
As in the HSCM, it is convenient to abuse the notations by introducing the expected degree and strength random variables $\kappa$ and $\sigma$ via
\begin{align}
  \kappa &= \kappa(\nu,\mu), \\
  \sigma &= \sigma(\nu,\mu),
\end{align}
thus changing latent random variables from $\nu,\mu$ to $\kappa,\sigma$. The joint distribution of the latter is given by the standard formula
\begin{equation}\label{eq:whscm-transformation-jacobian}
  \rho_{\kappa,\sigma}(\kappa,\sigma)=\rho_{\nu,\mu}[\nu(\kappa,\sigma),\mu(\kappa,\sigma)]\left|\frac{\partial(\nu,\mu)}{\partial(\kappa,\sigma)}\right|,
\end{equation}
where $\nu(\kappa,\sigma),\mu(\kappa,\sigma)$ are the inverse functions of $\kappa(\nu,\mu),\sigma(\nu,\mu)$, and $\left|\partial(\nu,\mu)/\partial(\kappa,\sigma)\right|$ is the absolute value of the determinant of the Jacobian:
\begin{equation}
  \left|\frac{\partial(\nu,\mu)}{\partial(\kappa,\sigma)}\right| = \left|\frac{\partial\nu}{\partial\kappa}\frac{\partial\mu}{\partial\sigma}-\frac{\partial\nu}{\partial\sigma}\frac{\partial\mu}{\partial\kappa}\right|.
\end{equation}

By the definition of $\kappa$ and $\sigma$, their joint distribution $\rho(\kappa,\sigma)$ is the joint distribution of expected degrees and strengths in the WHSCM, the analogy of the joint expected degree sequence $\kappa_i, \sigma_i$~(\ref{eq:wscm-kappa-i},\ref{eq:wscm-sigma-i})---joint via node index $i$---in the WSCM. Therefore, the WHSCM analogy of the WSCM constraints~(\ref{eq:wscm-kappa-constraints},\ref{eq:wscm-sigma-constraints}) is
\begin{equation}\label{eq:whscm-constraints}
  \rho(\kappa,\sigma)=\rho^\ast(\kappa,\sigma),
\end{equation}
where $\rho^\ast(\kappa,\sigma)$ is any desired joint distribution of expected degrees and strengths.

The marginal distributions $\rho(\kappa)$ and $\rho(\sigma)$ of the joint distribution $\rho(\kappa,\sigma)$ are the distributions of expected degrees and strengths in the ensemble. Therefore the expected average degree and strengths in the model are given by
\begin{align}
    \bar{k} &= \bar{\kappa}=\int \kappa\rho(\kappa)\,d\kappa = \iint \kappa(\nu, \mu) \rho(\nu, \mu) \, d\nu \, d\mu,\label{eq:whscm-kbar}\\
    \bar{s} &= \bar{\sigma}=\int \sigma\rho(\sigma)\,d\sigma = \iint \sigma(\nu, \mu) \rho(\nu, \mu) \, d\nu \, d\mu.\label{eq:whscm-sbar}
\end{align}

The joint distribution $\rho(\kappa,\sigma)$ of expected degrees $\kappa$ and strengths $\sigma$ defines the joint distribution of actual degrees $k$ and strengths $s$ via
\begin{equation}\label{eq:whscm-general-joint-degree-strength-distribution}
    P(k,s) = \iint P(k,s | \kappa, \sigma) \rho(\kappa, \sigma) \, d\kappa \, d\sigma.
\end{equation}
Unfortunately, the conditional joint distribution $P(k,s | \kappa, \sigma)$ of degrees and strengths $k,s$ of nodes of a given expected degree and strength $\kappa,\sigma$ is in general unknown. It is not even known, in general, what the closed form expression is for the conditional distribution $P(s|\sigma)$ of strengths $s$ of nodes of a given expected strength $\sigma$,
which appears in the expression for the strength distribution
\begin{equation}\label{eq:P(s)-WHSCM}
    P(s) = \int P(s | \sigma) \rho(\sigma) \, d\sigma.
\end{equation}
In view of Eq.~\eqref{eq:wscm-weight-sampling}, $P(s|\sigma)$ is the distribution of a sum of exponential random variables with different \emph{random} rates. The best what is known about such distributions---that are called \emph{hypoexponential distributions}---are some bounds on their tails, but only for fixed, not random rates~\cite{janson2018bounds}.
These distributions $P(s|\sigma)$ are definitely not as simple and well-studied as Poisson distributions $P(k|\kappa)$, so that very little appears to be known about mixtures of the former {\it a la} in~\eqref{eq:P(s)-WHSCM}. However, it is known~\cite{boguna2003class} that for sparse graphs the conditional distribution $P(k|\kappa)$ of degrees $k$ of nodes with a given expected degree $\kappa$ in the WHSCM is still Poisson, as in the HSCM, so that Eq.~(\ref{eq:hscm-P(k)}) holds in the WHSCM as well.

In Appendix~\ref{sec:proof}, we generalize the notions of sparse graphons and their entropy to weighted networks. This generalization allows us to discuss how to extend the HSCM maximum entropy proof to the WHSCM case, thus showing that the entropy of graphs in the WHSCM ensemble defined above is maximized across all graphs whose joint distribution of strengths and degrees converges to a given joint distribution.

The main focus of the rest of the paper is to identify the latent parameter distribution $\rho(\nu,\mu)$ that leads to joint degree-strength distributions $P(k,s)$ observed in real-world weighted networks.
In what follows, we first formulate in the next section this real-world-inspired form of $P(k,s)$ that we want our specific version of the general WHSCM to reproduce, and then derive the $\rho(\nu,\mu)$ that leads to this $P(k,s)$.

\section{Specific WHSCM requirements}\label{sec:requirements}

According to our general definition of the WHSCM model in the previous section, a specific version of the model is fixed by a particular choice of the joint degree-strength distribution $P(k,s)$. Here we document a specific set of properties of this joint distribution, dictated by the properties of many real-world weighted networks~\cite{barrat2004architecture,bagler2008analysis,barthelemy2011spatial,popovic2012geometric,allard2017geometric}, that we want our specific version of the general WHSCM introduced above to reproduce.

\textit{First}, we require the degree distribution $P(k)$, a marginal of $P(k,s)$, to be a power law
\begin{equation}
  P(k) \sim k^{-\gamma}
\end{equation}
with $\gamma > 2$. Here by ``power law'' and the ``$\sim$'' sign, we mean that $P(k)$ is a \textit{regularly varying} distribution~\cite{voitalov2019scale} which is a distribution whose complementary CDF satisfies
\begin{equation}
  \bar{F}(k) = \ell(k) k^{-(\gamma - 1)}, 
\end{equation}
where $\ell(k)$ is a \textit{slowly varying function}. Our power laws will be Pareto-mixed Poisson distributions whose $\ell(k)$s converge to constants, $\lim_{k\to\infty}\ell(k)=c$.

\textit{Second}, we require the strength of nodes to grow super-linearly with their degrees, as observed in many real weighted networks~\cite{barrat2004architecture,popovic2012geometric,allard2017geometric}. This observation is often expressed as
\begin{equation}
  \bar{s}(k) \sim k^{\eta},
\end{equation}
where $\eta \geq 1$ and $\bar{s}(k)$ is the average strength of nodes of degree $k$.
We interpret this relation to mean that
\begin{equation}
  \lim_{k \to \infty} \frac{\bar{s}(k)}{k^{\eta}} = s_0
\end{equation}
for some constant $s_0$.

If the distributions $P(s|k)$ of strengths $s$ of nodes of degree $k$ are concentrated around their expected values, and the degree distribution $P(k)$ is a power law with exponent $\gamma$, then the resulting strength distribution $P(s)$ is a power law with exponent $\delta$:
\begin{align}
  P(s) &\sim s^{-\delta}, \\
  \delta &= 1 + \frac{\gamma - 1}{\eta}.
\end{align}
If $\eta > \gamma - 1$, then the strength distribution $P(s)$ has exponent $\delta < 2$, meaning that for such combinations of $\gamma$ and $\eta$, the strength distribution $P(s)$ has an infinite first moment, so that the average strength $\bar{s}$ diverges. We do not want to exclude this possibility from our model, since such combinations of the values of $\gamma$ and $\eta$ can be found in real networks~\cite{allard2017geometric}.

\textit{Third}, we want our model to produce networks whose average degree $\bar{k}$, given by Eq.~\eqref{eq:whscm-kbar}, is independent of the network size $n$. A model satisfying this condition is said to produce \textit{sparse} networks.

To simplify the problem significantly, we next observe that if the actual degrees $k$ and strengths $s$ are concentrated around their expected values $\kappa$ and $\sigma$---and this is indeed the case as we show in Appendix~\ref{sec:app-concentrations}---then the requirements discussed above can be formulated in terms of $\kappa,\sigma$ instead of $k,s$.

We are thus looking for a model in which the distribution $\rho(\kappa)$ of expected degrees $\kappa$ follows a power law with exponent $\gamma > 2$,
\begin{equation}\label{eq:whscm-pl-kappa}
  \rho(\kappa) \sim \kappa^{-\gamma},
\end{equation}
and the expected strengths $\sigma$ grow super-linearly with expected degrees $\kappa$. For further simplicity, we want expected strengths and degrees to be deterministically related via
\begin{equation}\label{eq:whscm_linear_relation_sigma_kappa}
  \sigma = \sigma_0 \kappa^{\eta},
\end{equation}
where $\sigma_0 > 0$. This choice instantly fixes the joint expected degree-strength distribution to
\begin{equation}\label{eq:whscm-pl-contraint}
    \rho(\kappa, \sigma) = \rho(\kappa) \delta(\sigma - \sigma_0 \kappa^{\eta}),
\end{equation}
where $\delta()$ is the Dirac delta function, while
the expected strengths are distributed as
\begin{align}
  \rho(\sigma) & \sim \sigma^{-\delta}, \text{ where}\\
  \delta & = 1 + \frac{\gamma - 1}{\eta}.
\end{align}

The parameters of a WHSCM model satisfying the discussed requirements are thus
\begin{equation}
  \gamma,\delta,\bar{k},\sigma_0.
\end{equation}
The reason for having $\sigma_0$ as a parameter instead of the more natural average strength $\bar{s}$, given by Eq.~\eqref{eq:whscm-sbar}, is that the latter is actually infinite if $\delta<2$ as discussed above. However, $\sigma_0$ is well defined even in this case, and controls the baseline of the scaling of strength as a function of degree. If $\delta>2$, the expected average strength $\bar{s}$ is finite and in one-to-one relation to $\sigma_0$ via
\begin{equation}
  \bar{s} = \sigma_0 \int \kappa^{\eta} \rho(\kappa) \, d\kappa.
\end{equation}

\section{Numerical experiments in search of a solution}\label{sec:mle}

We are to find the latent parameter distribution $\rho(\nu,\mu)$ that yields the distribution $\rho(\kappa,\sigma)$ of expected degrees and strengths that we want in Eq.~\eqref{eq:whscm-pl-contraint}. Unfortunately, the accomplishment of this task using brute force does not appear possible since it involves solving a system of nonlinear integral equations, as we will see below. Therefore, in this section, we retreat to numeric experiments and describe a workaround that relies on getting some experimental hints from the WSCM.

First, for convenience and consistency with the power-law HSCM described in Sec.~\ref{sec:background}, we change variables from $\nu$ to $\lambda$ via 
\begin{equation}\label{eq:lambda-definition}
    \lambda = e^{R - \nu}.
\end{equation}
The support of $\nu$ is $(-\infty,R]$, so that the support of $\lambda$ is $[1,\infty)$. We will see that $R$ is the parameter that controls the average degree $\bar{k}$ and constant $\sigma_0$, and that it grows logarithmically with $n$, Appendix~\ref{sec:app-R-a-scaling}. With this change of variables, the connection probability and expected weight change from (\ref{eq:wscm-conn-prob},\ref{eq:whscm-omega}) to
\begin{align}
    p(\lambda_i, \mu_i, \lambda_j, \mu_j) &= \frac{1}{1 + e^{2R}\cdot\frac{\mu_i + \mu_j}{\lambda_i \lambda_j}},\label{eq:whscm-conn-prob}\\
    \omega(\lambda_i, \mu_i, \lambda_j, \mu_j) &= \frac{1}{1 + e^{2R}\cdot\frac{\mu_i + \mu_j}{\lambda_i \lambda_j}} \cdot \frac{1}{\mu_i + \mu_j},\label{eq:whscm-exp-weight}
\end{align}
while the expected degree and strength as functions of the latent variables change from (\ref{eq:whscm-bar(k)},\ref{eq:whscm-bar(s)}) to
\begin{align}
    \kappa(\lambda, \mu) &= n \int\limits_{1}^{\infty}d\lambda^{\prime} \int\limits_{0}^{\infty} d\mu^{\prime}\,\frac{ \rho(\lambda^{\prime}, \mu^{\prime}) }{1 + e^{2R} \cdot \frac{\mu + \mu^{\prime}}{\lambda \lambda^{\prime}}},\label{eq:whscm-exp-degree}\\
    \sigma(\lambda, \mu) &= n \int\limits_{1}^{\infty} d\lambda^{\prime} \int\limits_{0}^{\infty} d\mu^{\prime}\,\frac{\rho(\lambda^{\prime}, \mu^{\prime})}{\left( 1 + e^{2R} \cdot \frac{\mu + \mu^{\prime}}{\lambda \lambda^{\prime}} \right) \Bigl( \mu + \mu^{\prime} \Bigr)}.\label{eq:whscm-exp-strength}
\end{align}
Observe that since weights are positive, the support of $\mu$ is $(0,\infty)$.

With this change of variables our task becomes to find $\rho(\lambda, \mu)$ producing $\kappa=\kappa(\lambda, \mu)$ and $\sigma=\sigma(\lambda, \mu)$ such that
\begin{enumerate}
  \item $\kappa$ is a power-law-distributed random variable~\eqref{eq:whscm-pl-kappa},
  \item $\sigma$ is a superlinear function of $\kappa$~\eqref{eq:whscm_linear_relation_sigma_kappa}.
\end{enumerate}
The brute-force solution of this task is the solution of the system of the two two-dimensional nonlinear integral equations~(\ref{eq:whscm-exp-degree},\ref{eq:whscm-exp-strength}) with respect to $\rho(\lambda, \mu)$, so that the resulting $\rho(\kappa, \sigma)$ is as required in~\eqref{eq:whscm-pl-contraint}. The required amount of brute force for this task is well beyond our analytical strength, so that we have to retreat to numeric investigations.

\subsection{WHSCM with power-law hidden-variable distributions}

The most reasonable choice of $\rho(\lambda,\mu)$ appears to be a clean power-law one. Indeed, as recalled in Sec.~\ref{ssec:max-ent-unweighted-background}, such a clean power-law choice of the distribution of latent parameters results in power-law degree distributions in the HSCM, so that one may expect the situation to be similar in the WHSCM. Unfortunately, this expectation is not correct as we show next.

\begin{figure}[htp]
 \centering
 \includegraphics[width=.49\textwidth]{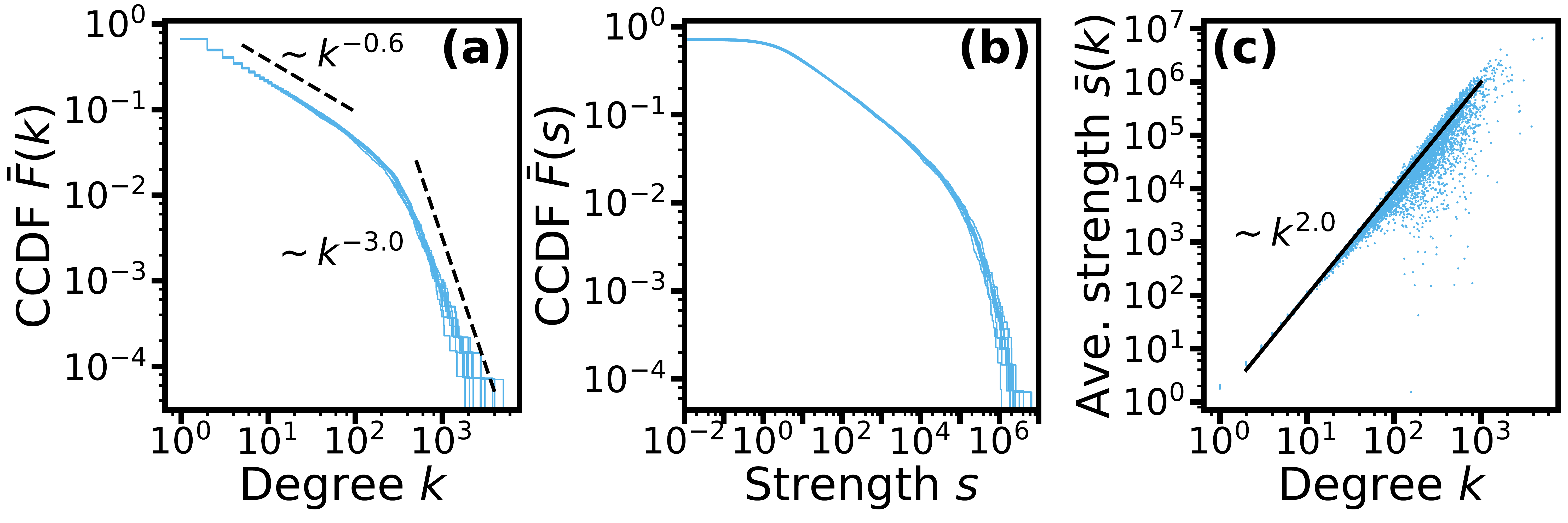}
  \caption{
  {\bf Basic structural properties of networks in the WHSCM with power-law hidden-variable distributions in  Eqs.~(\ref{eq:whscm-single-exp-lambda}-\ref{eq:whscm-single-exp-mu}).}
  Ten networks of size $n = 20,000$ are generated with $\alpha = 2.5, \beta = 2.0, a = 1.0, R = 6.0$. The resulting average degree is about~$20$. Panels \textbf{(a,b)} show the complementary cumulative distribution functions (CCDF) of degrees and strengths, while panel \textbf{(c)} shows the average strengths of nodes of degree~$k$.
  }
  \label{fig:simple-pl-whscm}
\end{figure}

To see this, let us set the distribution $\rho(\lambda)$ to be a pure power law, i.e., the Pareto distribution,
\begin{equation}\label{eq:whscm-single-exp-lambda}
    \rho(\lambda) = (\alpha - 1) \lambda^{-\alpha},\, \alpha > 2,\, \lambda>1,
\end{equation}
and couple $\lambda$ and $\mu$ deterministically via
\begin{align}\label{eq:whscm-single-exp-joint}
    \rho(\lambda, \mu) &= \rho(\lambda) \delta(\mu - f(\lambda)),\\
    f(\lambda) &= a \lambda^{-\beta},\,\beta \geq 0,\label{eq:whscm-single-exp-mu}
\end{align}
where $\delta()$ is the Dirac delta function, so that $\rho(\mu) \sim \mu^{(\alpha-1)/ \beta - 1}$ is another Pareto if $\beta > \alpha - 1$. We then generate random networks in this version of the WHSCM, and report their basic structural properties in Fig.~\ref{fig:simple-pl-whscm}.

We see that contrary to our expectations, the degree and strength distributions do not look like clean power laws, and that the degree distribution appears to be a double power law. Appendix~\ref{sec:app-single-exponent-whscm} contains some analytical arguments explaining this behavior.

\subsection{Experimental hints from the WSCM}

Given that the simple Pareto choice of the joint probability distribution $\rho(\lambda, \mu)$ does not lead to weighted networks with desired properties, we devise a workaround, looking for an ansatz for $\rho(\lambda, \mu)$ based on hints from the WSCM. Specifically, we attempt to ``reverse engineer'' the distribution $\rho(\lambda, \mu)$ by inferring the values of variables $\lambda_i, \mu_i$ in the WSCM for synthetically generated degree and strength sequences that satisfy our desired WHSCM constraints in Sec.~\ref{sec:requirements}.

The inference is done using maximum likelihood estimation (MLE) by finding a set of values $\lambda_i,\mu_i$ that maximize the log-likelihood $L = \log{P(G)}$ of a weighted graph $G$ in the WSCM. The probability $P(G)$ of $G$ in the WSCM ensemble is given by Eq.~\eqref{eq:wscm-gibbs}. The partition function $Z$ in that equation is calculated in~\cite{gabrielli2019grand} to yield the following explicit expression for $P(G)$:
\begin{equation}
\begin{split}
    P(G) = &\prod\limits_{(i,j) \in \mathcal{P}} \frac{e^{-(\nu_i + \nu_j + \log{(\mu_i + \mu_j)}) a_{ij}}}{1 + e^{-(\nu_i + \nu_j + \log{(\mu_i + \mu_j)})}} \cdot\\ 
                                 \cdot &\prod\limits_{(i,j) \in \mathcal{E}} \left( \mu_i + \mu_j \right) e^{-(\mu_i + \mu_j) w_{ij}},
\end{split}
\end{equation}
where $a_{ij}$ and $w_{ij}$ denote the entries of the adjacency and weight matrices of $G$, and $\mathcal{P}$ and $\mathcal{E}$ denote the sets of node pairs and connected node pairs in $G$. Using
\begin{align}
    \sum\limits_{(i,j) \in \mathcal{P}} (\nu_i + \nu_j) a_{ij} &= \sum\limits_{i = 1}^{n} k_i\nu_i,\\
    \sum\limits_{(i,j) \in \mathcal{P}} (\mu_i + \mu_j) w_{ij} &= \sum\limits_{i = 1}^{n} s_i\mu_i,
\end{align}
where $k_i,s_i$ are the degree and strength of node $i$ in graph $G$, the expression for the log-likelihood simplifies to
\begin{equation}\label{eq:prob-joint-wscm}
\begin{split}
    \log P(G) &= \sum\limits_{i=1}^{n} \Bigl[ k_i \nu_i + s_i \mu_i \Bigr] -\\
              &- \sum\limits_{(i,j) \in \mathcal{P}} \log \Bigl[ 1 + \frac{1}{e^{\nu_i + \nu_j} (\mu_i + \mu_j)} \Bigr].
\end{split}
\end{equation}
Observe that $\log{P(G)}$ depends on graph $G$ only via its degree $\boldsymbol{k}$ and strength $\boldsymbol{s}$ sequences, because they are the sufficient statistics in this grand-canonical ensemble. Furthermore, the probability $P(\boldsymbol{k},\boldsymbol{s}|\boldsymbol{\kappa},\boldsymbol{\sigma})$ of the degree-strengths sequence $\boldsymbol{k},\boldsymbol{s}$ in the WSCM ensemble defined by the expected degree-strength sequence $\boldsymbol{\kappa},\boldsymbol{\sigma}$ is at its maximum for $\boldsymbol{k}=\boldsymbol{\kappa}$ and $\boldsymbol{s}=\boldsymbol{\sigma}$~\cite{squartini2011analytical,cimini2019statistical}. Therefore, to execute our MLE program, all we have to do is to generate synthetic degree-strength sequences satisfying the requirements in Sec.~\ref{sec:requirements}, and then feed them to the MLE applied to the $\log{P(G)}$ above.

We do so as follows: for $i=1,\ldots,n$, we sample real random numbers $x_{i}$ from the Pareto distribution with exponent $\gamma > 2$ and mean $\bar{x}=10$, and then round them to the closest integers to yield degrees sequences $k_{i} = [x_{i}]$.
The exponent $\gamma$ and mean $\bar{x}$ of the Pareto distribution determine the $x_{\min}=\bar{x}(\gamma-2)/(\gamma-1)$ of its support $[x_{\min},\infty)$, so that upon this rounding the $k_{\min}$ degree is statistically different from the other degrees, but this difference has no effect on the tail exponent of the resulting distribution of $k_i$s, which is always guaranteed to be the same~$\gamma$~\cite{voitalov2019scale}.
The strengths are then set to $s_i = \sigma_0 k_i^{\eta}$ for some $\sigma_0>0$ and $\eta \geq 1$. The obtained sequences of degrees $\{k_{1},\ldots,k_{n} \}$ and strengths $\{s_{1},\ldots,s_{n} \}$ are then supplied to the MLE inference of the parameters $\{\nu_1,\ldots,\nu_n\}$ and $\{\mu_1,\ldots,\mu_n\}$ by maximizing the $\log{P(G)}$ in Eq.~\eqref{eq:prob-joint-wscm} which is done using the simulated annealing algorithm available in the PaGMO package~\cite{francesco-biscani-2019-2529931}. The inferred $\{\nu_1,\ldots,\nu_n\}$ are then mapped to $\{\lambda_1,\ldots,\lambda_n\}$ via~\eqref{eq:lambda-definition} with $R=\max_i\nu_i$. Another way to find $\lambda_i, \mu_i$ is to solve numerically the system of $2n$ equations in Eqs.~(\ref{eq:wscm-kappa-constraints},\ref{eq:wscm-sigma-constraints}). As noted in~\cite{mastrandrea2014enhanced}, this way leads to the same results. In our experiments, however, we find that the MLE approach produces more numerically stable results, so that we use this approach instead.

\begin{figure*}
 \includegraphics[width=.85\textwidth]{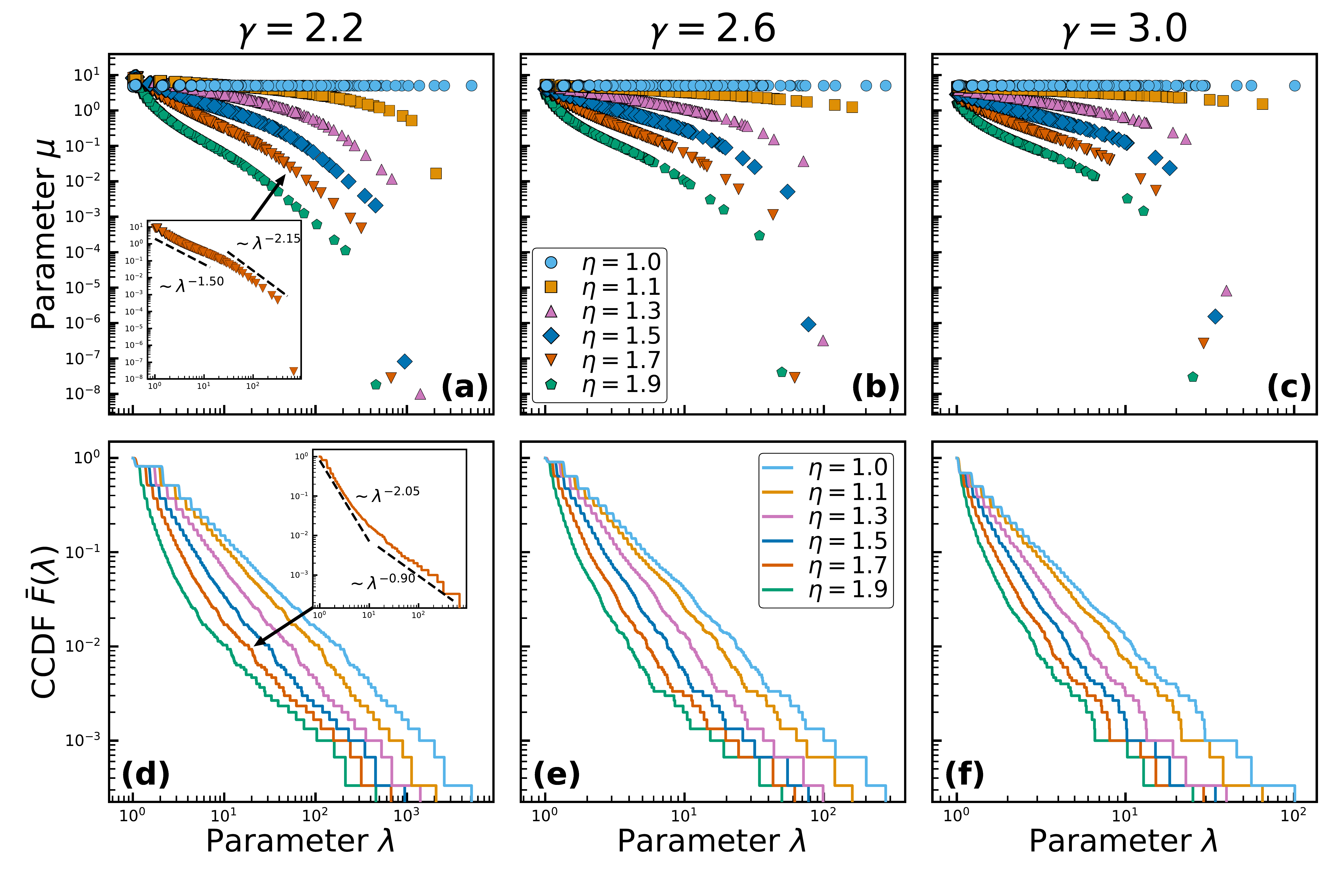}
  \caption{
  {\bf Maximum likelihood estimation of the latent parameters in the WSCM with power-law degree and strength sequences.}
  Degree sequences of length $n = 3,000$ are sampled from the Pareto distribution with varying power-law exponent $\gamma$, and then rounded to the closest integer. The corresponding strengths are set to $s_i = \sigma_0 k_i^{\eta}$. The expected average degree is set to $\bar{k} = 10$ and $\sigma_0 = 0.1$. Panels \textbf{(a), (b), (c)} show scatter plots of the inferred parameters $\lambda, \mu$ for varying values of $\eta$ and $\gamma = 2.2, 2.6, 3.0$. The inset in panel \textbf{(a)} shows the $\lambda$-$\mu$ scatter plot for $\eta = 1.7$, $\gamma = 2.2$ along with the power-law fit lines for small-$\lambda$ and large-$\lambda$ regions. Panels \textbf{(d), (e), (f)} show the complementary CDFs of the $\lambda$ parameter for varying values of $\eta$ and $\gamma = 2.2, 2.6, 3.0$. The inset in panel \textbf{(d)} shows the complementary CDF of the $\lambda$ parameter for $\eta = 1.7$, $\gamma = 2.2$ along with the power-law fit lines for small-$\lambda$ and large-$\lambda$ regions.
  }
  \label{fig:mle-figure}
\end{figure*}

The obtained sequences $\{\lambda_1,\ldots,\lambda_n\}$ and $\{\mu_1,\ldots,\mu_n\}$ for different values of $\gamma$ and $\eta$ are shown in Fig.~\ref{fig:mle-figure}. From this figure, we extract several hints suggesting a possible shape of the joint distribution of latent variables $\rho(\lambda, \mu)$:
\begin{enumerate}
  \item the variable $\mu$ can be directly coupled to the variable $\lambda$ via some function $f(\lambda)$ as indicated by the $\lambda$-$\mu$ scatter plots;
  \item the visual inspection of the $\lambda$-$\mu$ scatter plots on the log-log scale suggests that this function scales approximately as a power-law for small and large values of $\lambda$, potentially with two different exponents, i.e., $f(\lambda) \sim \lambda^{-\beta_1}$ when $\lambda \rightarrow 1$, and $f(\lambda) \sim \lambda^{-\beta_2}$ when $\lambda \gg 1$;
  \item the complementary CDF $\bar{F}(\lambda)$ behaves approximately as a power-law for small and large values of $\lambda$, potentially with two different exponents $\alpha_1$ and $\alpha_2$.
\end{enumerate}
 
\section{Weighted hypersoft configuration model with power-law degree and strength distributions}\label{sec:whscm}

Here we rely on the observations at the end of the previous section to specify a particular version of the WHSCM model that satisfies the requirements in Sec.~\ref{sec:requirements}. These observations instruct us to set the joint distribution of latent parameters $\lambda,\mu$ to
\begin{equation}\label{eq:whscm-joint-prob}
    \rho(\lambda, \mu) = \rho(\lambda) \delta(\mu - f(\lambda)),
\end{equation}
where $\delta()$ is the Dirac delta function. This setting fixes the WHSCM-definitive distribution $\rho(\nu,\mu)$ via the change of variables~\eqref{eq:lambda-definition}, but $\rho(\lambda)$ and $f(\lambda)$ are still to be specified.

We specify them, again following the hints from the end of the previous section, as follows. The distribution of $\lambda$ is set to
\begin{equation}\label{eq:whscm-rho-lambda}
    \rho(\lambda) = 
    \begin{cases}
         A_{1} \lambda^{-\alpha_1},\,\, 1 < \lambda \leq \lambda_c, \\
         A_{2} \lambda^{-\alpha_2},\,\, \lambda_c < \lambda < \infty,
    \end{cases}
\end{equation}
where $\alpha_1 > 1$, $\alpha_2 > 1$, and $\lambda_c$ is a crossover point between the two power laws with exponents $\alpha_1$ and $\alpha_2$, whereas $A_1$, $A_2$ are the normalization constants given by
\begin{align}
  A_{1} &= \frac{\left( \alpha_1 - 1 \right) \left( \alpha_2 - 1 \right)}{\lambda_{c}^{1 - \alpha_1} \left( \alpha_1 - \alpha_2 \right) + (\alpha_2 - 1)}, \\
  A_{2} &= A_{1} \lambda_{c}^{\alpha_2 - \alpha_1},
\end{align}
while the function $f(\lambda)$ is set to
\begin{equation}\label{eq:whscm-f-lambda}
    f(\lambda) = 
    \begin{cases}
        a \lambda^{-\beta_1},\,\, 1 < \lambda \leq \lambda_c, \\
        a \lambda_{c}^{\beta_2 - \beta_1} \lambda^{-\beta_2},\,\, \lambda_c < \lambda < \infty,
    \end{cases}
\end{equation}
where $a > 0$, $\beta_1 \geq 0$, $\beta_2 \geq 0$. With these settings, our model is fully specified by the following set of parameters:
\begin{enumerate}
  \item exponents $\alpha_1, \alpha_2, \beta_1, \beta_2$, and
  \item parameters $R$ and $a$,
\end{enumerate}
that we move on to specify below. The double power law crossover point $\lambda_c$ may also appear as a free parameter, but we set it to be a function of other parameters as follows.

Recall that the connection probability $p(\lambda_i, \mu_i, \lambda_j, \mu_j)$ in the model is given by Eq.~\eqref{eq:whscm-conn-prob}. The crossover point $\lambda_c$ is selected in such a way that for $\lambda_i,\lambda_j$ below $\lambda_c$ this connection probability can be approximated by dropping the $1$ in the denominator, analogous to the classical limit approximation of the Fermi-Dirac distribution function in statistical physics~\cite{van2018sparse}. This means that the constant $\lambda_c$ defines the point where the term $1$ in the denominator in \eqref{eq:whscm-conn-prob} is comparable to the other term there. Therefore, we define $\lambda_c$ to be the value of $\lambda_i, \lambda_j$ such that $e^{2R} \times \frac{\mu_i + \mu_j}{\lambda_i \lambda_j} = 1$, so that
\begin{equation}\label{eq:whscm-lambda-c}
    \lambda_c = \left( 2 a e^{2R} \right)^{\frac{1}{2 + \beta_1}}.
\end{equation}

We show in Appendix~\ref{sec:app-scalings} that with the settings above, the expected degree of a node with latent parameter $\lambda$ can be written as
\begin{equation}\label{eq:whscm-exp-degree-2}
    \kappa(\lambda) =
    \begin{cases}
        I_1 (\lambda) + I_2 (\lambda),\,\, 1 < \lambda \leq \lambda_c, \\
        I_3 (\lambda) + I_4 (\lambda),\,\, \lambda_c < \lambda < \infty,
    \end{cases}
\end{equation}
where the integrals $I_1$-$I_4$ are explicitly defined in Eqs.~(\ref{eq:whscm-i1}-\ref{eq:whscm-i4}). Similarly, the expected strength of a node with latent parameter $\lambda$ can be written as
\begin{equation}\label{eq:whscm-exp-strength-2}
    \sigma(\lambda) =
    \begin{cases}
        I_5 (\lambda) + I_6 (\lambda),\,\, 1 < \lambda \leq \lambda_c, \\
        I_7 (\lambda) + I_8 (\lambda),\,\, \lambda_c < \lambda < \infty,
    \end{cases}
\end{equation}
where the integrals $I_5$-$I_8$ are explicitly defined in Eqs.~(\ref{eq:whscm-i5}-\ref{eq:whscm-i8}). We also show in Appendix~\ref{sec:app-scalings} that the functions $\kappa(\lambda)$ and $\sigma(\lambda)$ can be well approximated by power laws
\begin{align}
    \kappa(\lambda) &\sim
    \begin{cases}
        \lambda^{\phi_1},\,\, \text{if } 1 < \lambda \leq \lambda_c, \\
        \lambda^{\phi_2},\,\, \text{if } \lambda_c < \lambda < \infty,
    \end{cases}
    \\
    \sigma(\lambda) &\sim
    \begin{cases}
        \lambda^{\chi_1},\,\, \text{if } 1 < \lambda \leq \lambda_c, \\
        \lambda^{\chi_2},\,\, \text{if } \lambda_c < \lambda < \infty,
    \end{cases}
\end{align}
where $\phi_1 \geq 1$, $0 < \phi_2 \leq 1$, $\chi_1 \geq \phi_1$, $\chi_2 \geq \phi_2$, and that with these approximations, the distributions of expected degrees and strengths do indeed exhibit power-law behavior:
\begin{align}
    \rho(\kappa) &\sim
    \begin{cases}
        \kappa^{-(1 + \frac{\alpha_1 - 1}{\phi_1})},\,\, \text{if } \kappa(1) < \kappa \leq \kappa(\lambda_c), \\
        \kappa^{-(1 + \frac{\alpha_2 - 1}{\phi_2})},\,\, \text{if } \kappa(\lambda_c) < \kappa < \infty,
    \end{cases}
    \label{eq:whscm-p-kappa}
    \\
    \rho(\sigma) &\sim
    \begin{cases}
        \sigma^{-(1 + \frac{\alpha_1 - 1}{\chi_1})},\,\, \text{if } \sigma(1) < \sigma \leq \sigma(\lambda_c), \\
        \sigma^{-(1 + \frac{\alpha_2 - 1}{\chi_2})},\,\, \text{if } \sigma(\lambda_c) < \sigma < \infty.
    \end{cases}
    \label{eq:whscm-p-sigma}
\end{align}

Now, if we know how the scaling exponents $\phi_1, \phi_2, \chi_1, \chi_2$ behave as functions of the model parameters, we can easily select those parameters to ensure that $\rho(\kappa) \sim \kappa^{-\gamma}$ for all values of $\lambda$, and that $\rho(\sigma) \sim \sigma^{-\delta}$ with $\delta = 1 + \frac{\gamma-1}{\eta}$ as required in Sec.~\ref{sec:requirements}.
In Appendix~\ref{sec:app-scalings} we find $\phi_1, \phi_2, \chi_1, \chi_2$ as functions of $\alpha_1, \alpha_2, \beta_1, \beta_2$, and show that the following nontrivial choice of the latter as functions of $\gamma,\eta$ produces the desired outcome:
\begin{align}
    \alpha_1 &= 1 + \eta (\gamma - 1), \label{eq:whscm-alpha1} \\
    \beta_1 &= \Bigl( \gamma - \frac{\gamma - 2}{\gamma} \Bigr) \Bigl( \eta - 1 \Bigr), \label{eq:whscm-beta1} \\
    \alpha_2 &= 1 + \frac{\alpha_1 - 1}{1 + \beta_1} \left[ 1 + (\gamma - 2) \left(1 - \frac{1}{\eta}\right) \right], \label{eq:whscm-alpha2} \\
    \beta_2 &=
    \begin{cases}
        0,\,\, &\text{if }\eta = 1, \\
        \alpha_2 - 1 + \frac{\eta (\alpha_2 - 1)}{\gamma - 1},\,\, &\text{if }\eta > 1.
    \end{cases} \label{eq:whscm-beta2}
\end{align}
The choice of the model parameters above is obtained using a series of approximations outlined in Appendix~\ref{sec:app-scalings}. We find in simulations that this choice produces networks with desired properties if $\gamma > 2$ and $\eta \leq 2$ as we will see below.

Finally, we have to express the $R$ and $a$ parameters as functions of the average degree $\bar{k}$ and $\sigma_0$.
In the $\lambda$ terms, the average degree is equal to
\begin{equation}\label{eq:whscm-average-k}
    \bar{k} = \int\limits_{1}^{\infty} \kappa(\lambda) \rho(\lambda)\,d\lambda.
\end{equation}
Both $R$ and $a$ appear in this integral. The second equation defining these two parameters is $\sigma = \sigma_0 \kappa^{\eta}$. For any given value of the latent parameter $\lambda = \lambda_{0}$, we can write
\begin{equation}\label{eq:whscm-sigma0}
    \sigma_0 = \frac{\sigma(\lambda_0)}{\left[\kappa(\lambda_0)\right]^{\eta}}.
\end{equation}
For the reasons explained in Appendix~\ref{sec:app-scalings}, we set the $\lambda_0$ to the value such that $\kappa(\lambda_0)=\bar{k}$. By solving numerically the system of Eqs.~(\ref{eq:whscm-average-k}, \ref{eq:whscm-sigma0}) with the $\lambda_0$ set to this value, we find the values of the parameters $R$ and $a$. With these solutions, the model is fully specified.

{\bf Weighted hypersoft configuration model with power-law degree and strength distributions.} To summarize, the graphs in the described power-law WHSCM can be generated using the following algorithm:
\begin{enumerate}
    \item{For a given set of input parameters, which are the network size $n$, the degree distribution power law exponent $\gamma > 2$, the strength-degree scaling exponent $\eta \geq 1$, the average degree $\bar{k}$, and the constant $\sigma_0$ controlling the baseline of the strength-degree scaling,}
    \item{Find the model parameters $\alpha_1, \alpha_2, \beta_1, \beta_2, R, a$ using Eqs.~(\ref{eq:whscm-alpha1}-\ref{eq:whscm-sigma0}).}
    \item{For each node $i = 1,\ldots, n$, sample $\lambda_i$ from the PDF in Eq.~\eqref{eq:whscm-rho-lambda}, and set $\mu_i=f(\lambda_i)$ according to Eq.~\eqref{eq:whscm-f-lambda}.}
    \item{For each node pair $(i,j)$, draw the link between them with probability given by Eq.~\eqref{eq:whscm-conn-prob}.}
    \item{For each node pair $(i,j)$ linked at the previous step, draw the weight of the link from the exponential distribution with rate $\mu_i + \mu_j$, Eq.~\eqref{eq:wscm-weight-sampling}.}
\end{enumerate}
We provide an implementation of this algorithm at the GitHub repository~\cite{githubcode}.

\section{Simulation results for synthetic networks}\label{sec:synthetic-nets}

Here we check in simulations that the specific version of the general WHSCM model documented at the end of the previous section produces networks satisfying the requirements in Sec.~\ref{sec:requirements}.

\begin{figure*}
 \centering
 \includegraphics[width=.8\textwidth]{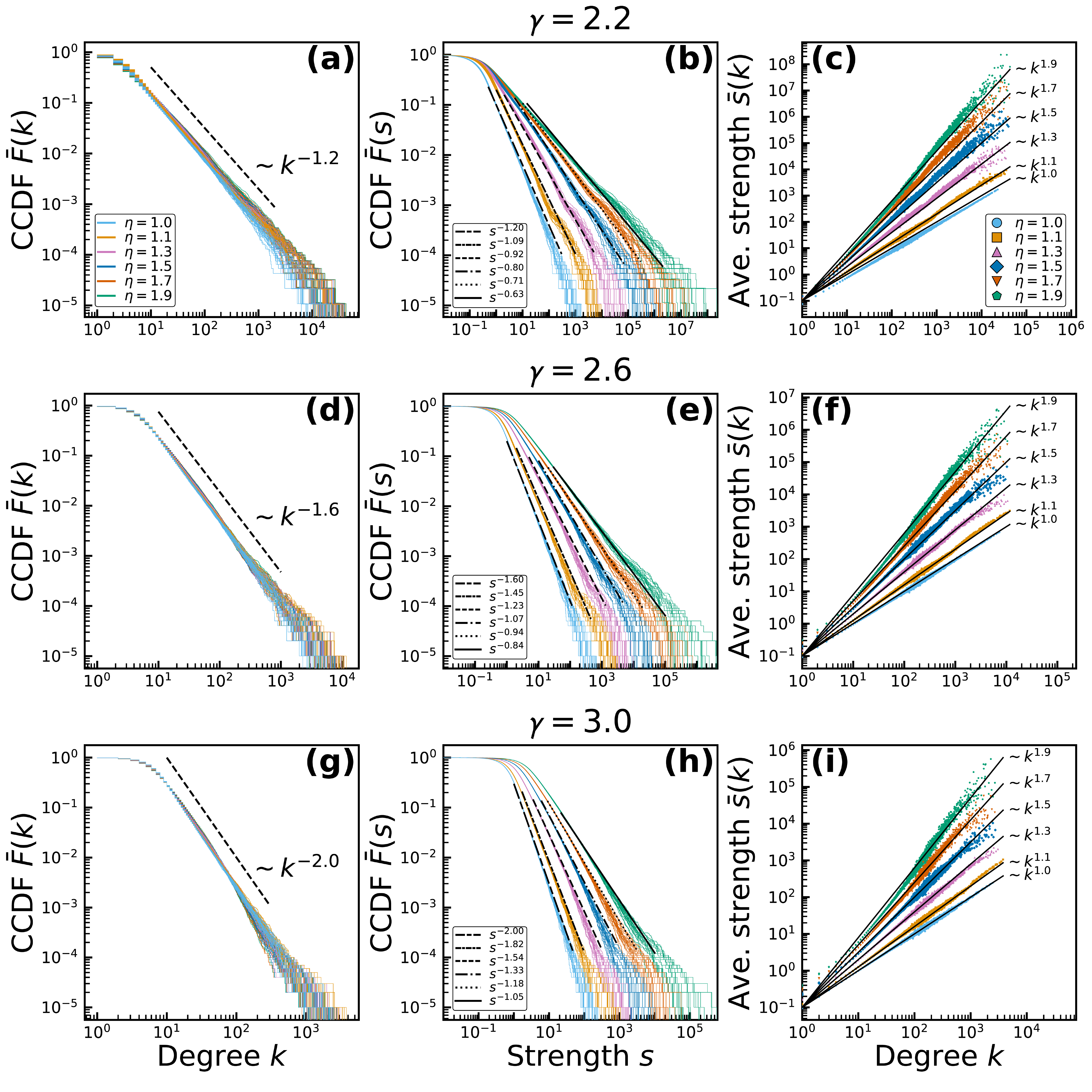}
  \caption{
  {\bf Structural properties of synthetic networks in the power-law WHSCM model defined in Sec.~\ref{sec:whscm}.}
  The parameters used to generate the networks are $n = 10^5$, $\bar{k} = 10$, $\sigma_0 = 0.1$, $\gamma \in \{2.2, 2.6, 3.0\}$, and $\eta \in \{1.0, 1.1, 1.3, 1.5, 1.7, 1.9\}$. For each combination of the parameters, $20$ network instances are generated. Panels \textbf{(a), (d)}, and \textbf{(g)} show the empirical complementary CDFs (CCDFs) of node degrees for $\gamma = 2.2, 2.6, 3.0$. Each panel shows $20$ degree CCDF curves corresponding to the $20$ network instances. The curves with the same value of exponent $\eta$ are of the same color. The black dashed lines show the imposed values of the power-law exponent $\gamma$. Panels \textbf{(b), (e)}, and \textbf{(h)} show the empirical CCDFs of node strengths for $\gamma = 2.2, 2.6, 3.0$. The black lines denote the power laws with the imposed values of $\delta - 1$ where the strength power-law exponent $\delta$ is related to $\gamma$ and $\eta$ via $\delta - 1 = (\gamma-1)/\eta$. Panels \textbf{(c), (f), (i)} show the strength-degree correlations for $\gamma = 2.2, 2.6, 3.0$. The average strength $\bar{s}(k)$ of nodes of degree $k$ is shown for all the $20$ network instances. The curves with the same $\eta$ are of the same color. The black lines show the functions $s(k) = \sigma_0 k^{\eta}$ with the imposed values of $\eta$. For clarity, the strength CCDFs are shown for $s \geq 10^{-2}$.
  }
  \label{fig:synthetic-fig}
\end{figure*}

\begin{figure*}
 \centering
 \includegraphics[width=.9\textwidth]{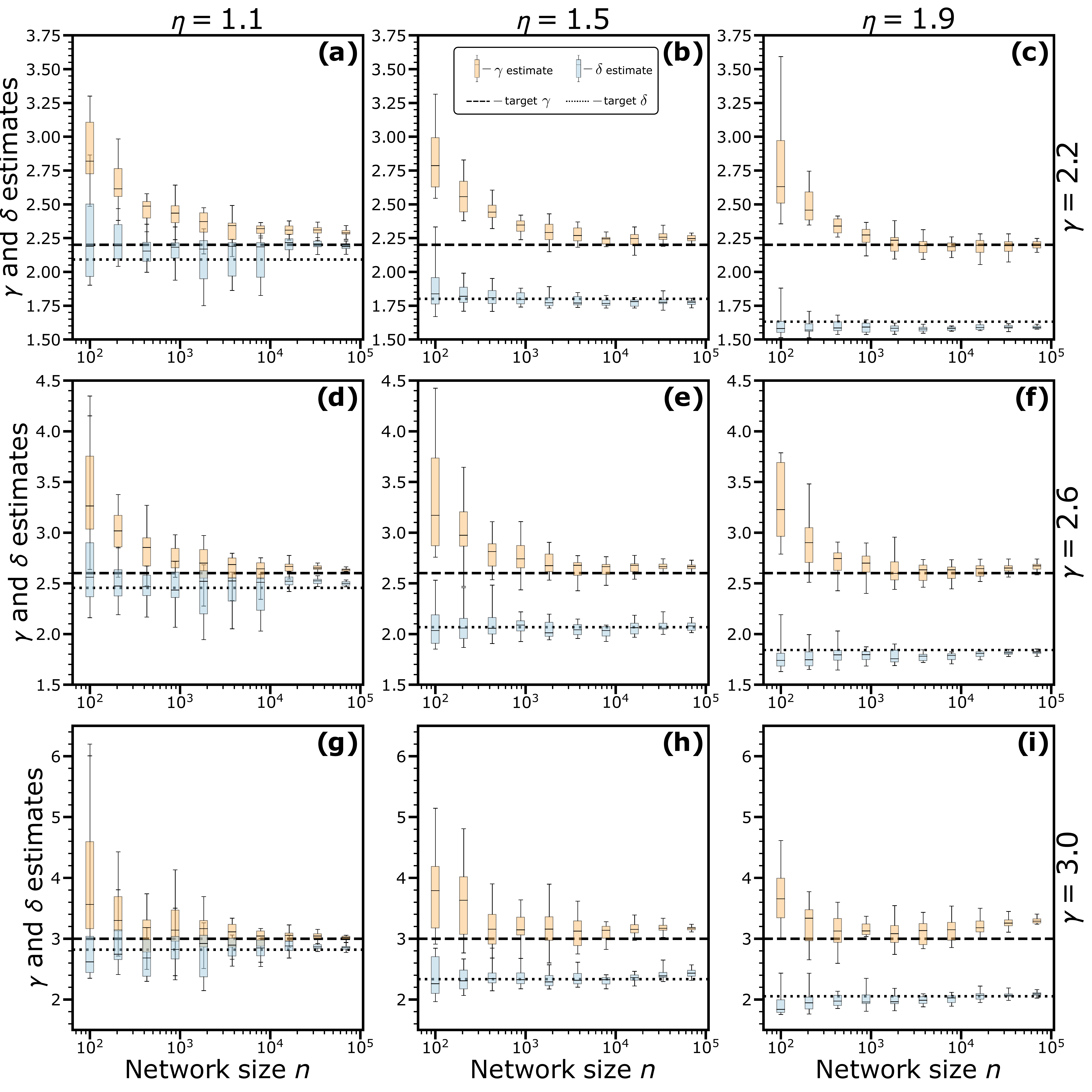}
  \caption{
  {\bf Inferred power-law exponents of the degree and strength distributions in synthetic WHSCM networks.}
  Synthetic networks of varying sizes $n \in [10^2, 10^5]$ are generated using the power-law WHSCM from Sec.~\ref{sec:whscm} with the shown matrix of values of parameters $\gamma,\eta$. All networks have $\bar{k} = 10$ and $\sigma_0 = 0.1$. For each combination of parameters and sizes~$n$, $20$ random networks are generated. The degree distribution power-law exponent $\gamma$ and the strength distribution power-law exponent $\delta$ are then inferred using the power-law exponent estimation software from~\cite{voitalov2019scale}.
  The box plots show the distributions of the inferred values of $\gamma,\delta$. The box plot settings are standard: the whiskers are the $5$-th and $95$-th percentiles, and the box boundaries are the $25$-th and $75$-th percentiles of the distributions; the black lines within the boxes are the medians. The dashed and dotted lines show the target values of $\gamma$ and $\delta$ used to generate the networks.
  }
  \label{fig:finite-n-exponent-estimates}
\end{figure*}

We first generate graphs of size $n = 10^5$ with the average degree $\bar{k} = 10$ and $\sigma_0 = 0.1$ for $\gamma \in \{2.2, 2.6, 3.0\}$ and $\eta \in \{1.0, 1.1, 1.3, 1.5, 1.7, 1.9\}$. For each combination of the parameters, $20$ graph instances are generated. The resulting empirical complementary CDFs of degrees $k$ and the average strengths $\bar{s} (k)$ of nodes of degree $k$ are shown in Fig.~\ref{fig:synthetic-fig}. We see that the generated networks indeed have power-law degree distributions with the prescribed values of exponent $\gamma$, and that the strength-degree correlations follow the super-linear law with the prescribed values of exponent $\eta$. Figure~\ref{fig:finite-n-exponent-estimates} quantifies this further, also showing the convergence of the inferred values of $\gamma$ and $\delta$ to their target values for the networks of growing sizes~$n$. Since the networks are random, so are their inferred $\gamma,\delta$. We see that the means of the distributions of $\gamma,\delta$ come closer to their target values as $n$ grows, while their variances decrease. The lack of an exact match in certain cases even for the largest considered network size can be attributed to that this size is still not sufficiently large, as well as to imperfections of the model and the exponent estimator. The estimation precision of the latter for a given $n$ is unknown~\cite{voitalov2019scale}.

To demonstrate how the degree-strength distributions, constrained in the maximum entropy manner, implicitly constrain some other network properties to some not exactly trivial values, we measure the \textit{weight disparity}~\cite{barthelemy2003spatial,barthelemy2005characterization,serrano2009extracting} in the generated networks. The weight disparity $Y_i$ is a quantity that characterizes the heterogeneity of weights of links incident to node $i$. It is defined as
\begin{equation}\label{eq:weight-disparity}
    Y_{i} =\sum\limits_{j \in \mathcal{N}_i} \left( \frac{w_{ij}}{s_i} \right)^2,
\end{equation}
where $\mathcal{N}_i$ denotes the set of neighbors of node $i$, $w_{ij}$ the weight of the link between nodes $i$ and $j$, and $s_i$ the strength of $i$. 
If most of the strength of node $i$ is concentrated in the weights of a few links incident to $i$, then $Y_i$ is close to $1$. If all the links incident to $i$ carry the same weight $s_i/k_i$, then $Y_i$ is equal to $1/k_i$. To characterize the local distribution of weights in a weighted network as a whole, the average weight disparity $\bar{Y}(k)$ of nodes of degree $k$ is often looked at~\cite{latora2017complex}. If $\bar{Y}(k) \sim 1/k$ for all degrees $k$, then the weights are homogeneously distributed among all links and all nodes. An average weight disparity function decaying slower than $1/k$, as observed in many real networks~\cite{serrano2009extracting,allard2017geometric}, indicates that the link weights are distributed more heterogeneously.

\begin{figure}
 \centering
 \includegraphics[width=.49\textwidth]{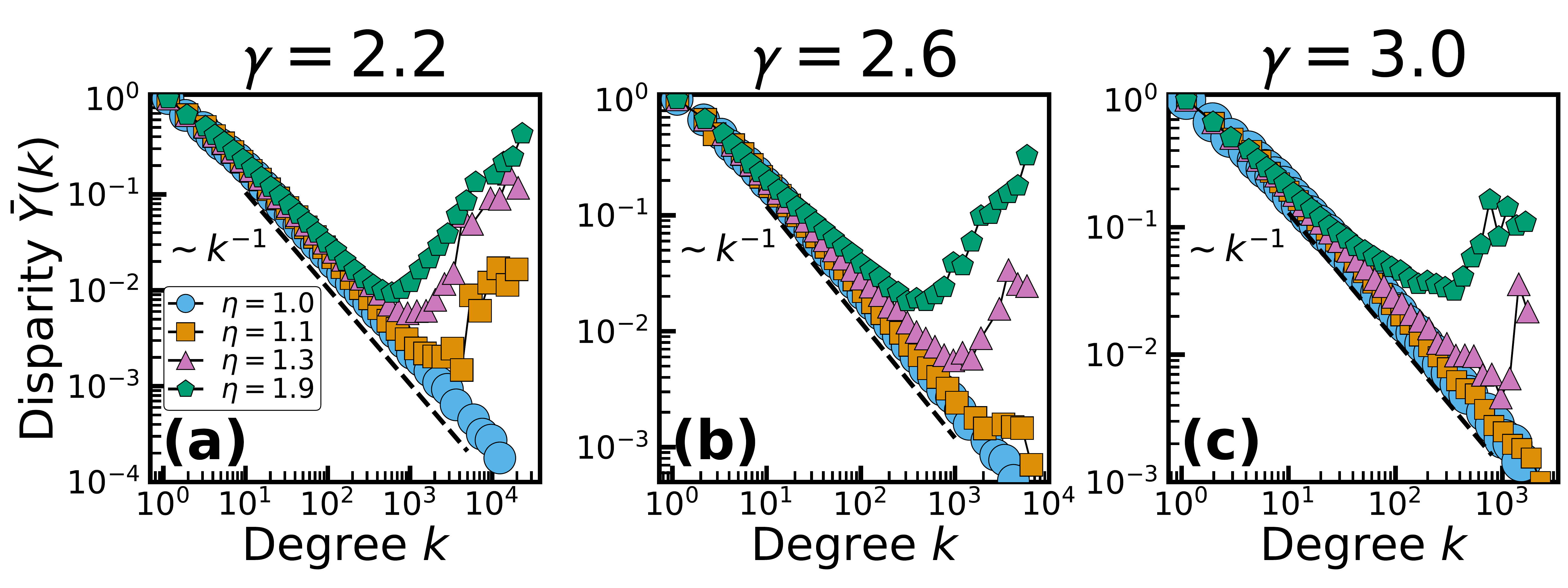}
  \caption{
  {\bf Weight disparity as a function of node degree in the power-law WHSCM.}
  The figure shows the log-binned average weight disparity $\bar{Y}(k)$ of nodes of degree $k$ in the power-law WHSCM networks with $\gamma = 2.2, 2.6, 3.0$ and $\eta = 1.0, 1.1, 1.3, 1.9$. The black dashed lines show the $1/k$ scaling. In all the generated networks, the other parameters are set to $n = 10^5$, $\bar{k} = 10$, and $\sigma_0 = 0.1$.
  }
  \label{fig:disparity-fig}
\end{figure}

In Fig.~\ref{fig:disparity-fig}, we show the average weight disparity functions  $\bar{Y}(k)$  in the generated WHSCM networks. We observe that the weight disparity behaves quite differently for different values of the scaling exponent $\eta$. For small values of $\eta$, the weight disparity as a function of degree scales roughly as $\bar{Y}(k) \sim 1/k$ in the lower to mid-range degree values. For larger values of $\gamma$ this behavior persists even for large degrees, indicating that weights are distributed homogeneously for nodes of both low and high degrees. However, if $\eta$ is larger, the high degree behavior of $\bar{Y}(k)$ changes entirely, and the function starts to increase with the degree $k$. This indicates that the higher the degree of a node, the more heterogeneous the distribution of weights among its links.

The intuition behind this effect is that the higher the exponent $\eta$, the larger the strengths of the high-degree nodes. In order for these nodes to satisfy their demanding strength constraints without disturbing the small strengths of low-degree nodes attached to them, they have to allocate increasingly heavier weights on their links to other high-degree nodes. This creates a weighted connectivity pattern where large portions of the strengths of high-degree nodes are distributed among the links interconnecting the high-degree, high-strength nodes. This weighted rich-club pattern is similar in spirit to the unweighted rich-club effect~\cite{colizza2006detecting,amaral2006lies,zhou2004rich}. It is important to reemphasize that this seemingly nontrivial effect is caused purely by simple constraints on low-order network properties---namely, by heterogeneous degree distributions and super-linear scalings of strengths with degrees. 

\section{Power-law WHSCM versus real-world networks}\label{sec:real-nets}

Finally, we demonstrate a use case scenario involving the power-law WHSCM to construct null model graphs for the following three real-world weighted networks:
\begin{enumerate}
  \item \textit{C.~elegans} metabolic network~\cite{jeong2000large,duch2005community,kunegis2013konect}, where nodes are metabolites and links indicate interactions between them, with the link weight representing the number of such interactions;
  \item computational geometry collaborations network~\cite{comp-geometry-data}, where nodes are authors publishing works on computational geometry and links between them represent co-authorships, with the link weight representing the number of co-authored works;
  \item Bible proper nouns network~\cite{bible-nouns-data,kunegis2013konect}, where nodes are proper nouns of places and names in the King James Version of the Bible and links between them indicate that a pair of nouns are mentioned in the same Bible verse, with link weights representing the number of such noun co-occurrences.
\end{enumerate}

To construct null model graphs for these networks, we first measure the WHSCM input parameters in these real networks. We find the $n$ and $\bar{k}$ parameters directly from the networks, the power-law exponent $\gamma$ is found using the package from Ref.~\cite{voitalov2019scale}, and the exponent $\eta$ and $\sigma_0$ are found by a linear fit of degrees and strengths in the log-log scale. The resulting values for the three networks are shown in Table~\ref{tab:real-nets-properties}.
\begin{table}[!h]
    \caption{The WHSCM parameters measured in the three real-world networks.
     }\label{tab:real-nets-properties}
    \begin{tabular}{p{0.23\textwidth} p{0.05\textwidth} p{0.04\textwidth} p{0.04\textwidth} p{0.035\textwidth} p{0.04\textwidth}}
        \hhline{======}
        \textbf{Network name} & $\boldsymbol{n}$ & $\boldsymbol{\overline{k}}$ & $\boldsymbol{\sigma_0}$ & $\boldsymbol{\gamma}$ & $\boldsymbol{\eta}$ \\
        \hline
        \rule{0pt}{3ex}\textit{C.\ elegans} metabolic~\cite{jeong2000large,duch2005community,kunegis2013konect} & $453$ & $8.94$ & $1.3$ & $2.5$ & $1.154$ \\
        Comp.\ geo.\ collab.~\cite{comp-geometry-data}  & $6,158$ & $3.86$ & $1.0$ & $2.6$ & $1.333$  \\
        Bible proper nouns~\cite{bible-nouns-data,kunegis2013konect} & $1,773$ & $10.30$ & $0.66$ & $3.1$ & $1.313$ \\
        \hhline{======}
    \end{tabular}
\end{table}

Using the parameters from Table~\ref{tab:real-nets-properties}, we generate ten synthetic WHSCM network instances for each of the three real networks, and compare the basic structural properties of the real networks and their WHSCM null model counterparts. Specifically, we look at the following properties:
\begin{enumerate}
  \item the complementary CDF (CCDF) of degrees, $\bar{F}(k)$;
  \item the average strength of nodes as a function of their degrees, $\bar{s}(k)$;
  \item the average weight disparity of nodes as a function of their degrees, $\bar{Y}(k)$;
  \item the average local clustering coefficient of nodes as a function of their degrees, $\bar{c}(k)$;
  \item the average weighted local clustering coefficient of nodes as a function of their strengths, $\bar{c}_w (s)$.
\end{enumerate}
The last, less familiar property is defined in~\cite{saramaki2007generalizations} for a node~$i$ as
\begin{equation}\label{eq:local-weighted-clustering}
    c_w^{(i)} = \frac{1}{k_i (k_i - 1)} \sum\limits_{(j,k) \in \mathcal{N}(i)} \left( \widetilde{w}_{ij} \widetilde{w}_{jk} \widetilde{w}_{ki} \right)^{1/3},
\end{equation}
where $(j,k)$ are all the pairs of neighbors $\mathcal{N}(i)$ of the node $i$, and $\widetilde{w}$ denotes the link weight {rescaled by the maximum weight} observed in the network. The function $\bar{c}_w (s)$ is the average of $c_w^{(i)}$ across all nodes $i$ whose strength is $s$. If weights are real, every node has a unique strength with high probability, so that this function is a scatter plot consisting of $n$ data points, one data point for every node.
\begin{figure*}
 \includegraphics[width=.99\textwidth]{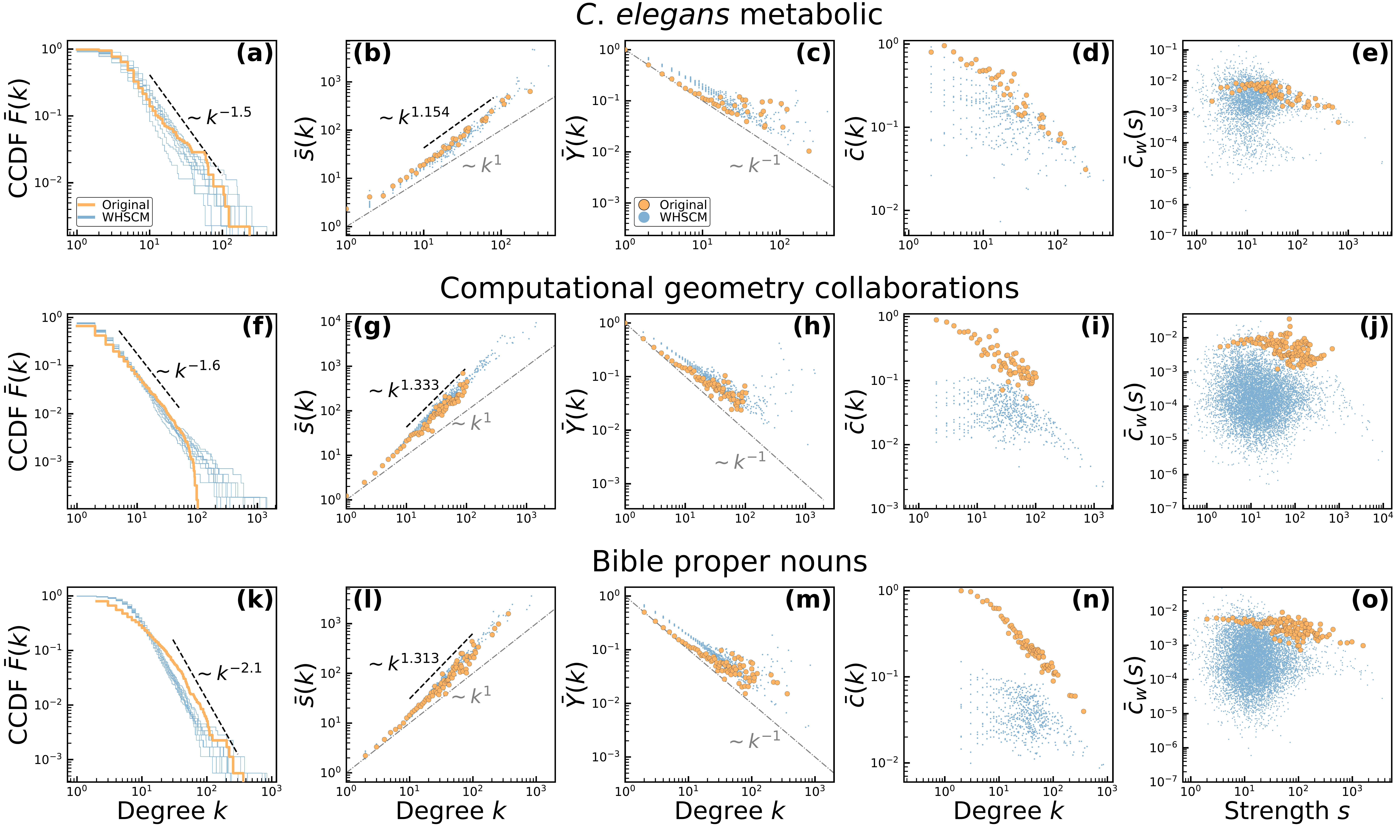}
  \caption{
  {\bf Real networks vs.\ their WHSCM counterparts.}
  Each row in the figure corresponds to one of the three real weighted networks described in Sec.~\ref{sec:real-nets}, as indicated by the network names at the top of each row. Each column in the figure shows different structural properties of the real networks: panels \textbf{(a), (f), (k)} show the degree CCDFs, with the dashed blacked lines indicating the pure power-law scalings with the exponents $\gamma$ listed in Table~\ref{tab:real-nets-properties}; panels \textbf{(b), (g)}, and \textbf{(l)} show the average strength as a function of node degree, $\bar{s}(k)$, with the black dashed lines indicating the pure scalings with the exponents $\eta$ shown in Table~\ref{tab:real-nets-properties}, and the gray dashed lines indicating the linear scaling; panels \textbf{(c), (h)}, and \textbf{(m)} show the average weight disparity, defined in Eq.~\eqref{eq:weight-disparity}, as a function of node degree, with the gray dashed lines indicating the $1/k$ scaling of the weight disparity corresponding to the perfectly homogeneous distribution of local weights; panels \textbf{(d), (i)}, and \textbf{(n)} show the average local clustering coefficient as a function of node degree; panels \textbf{(e), (j)}, and \textbf{(o)} show the average weighted local clustering coefficient, defined in Eq.~\eqref{eq:local-weighted-clustering}, as a function of node strength. The data for the real networks are shown in the orange color, while the data for the synthetic WHSCM networks are in blue. For each real networks, ten random WHSCM graphs are generated using the power-law WHSCM with the parameters listed in Table~\ref{tab:real-nets-properties}. 
  }
  \label{fig:real-nets}
\end{figure*}

Juxtaposing these properties in the considered real networks against their WHSCM counterparts in Fig.~\ref{fig:real-nets},  we observe that the properties fixed by the WHSCM are well-captured by the null model graphs, as expected. Moreover, even though the weight disparity is not explicitly constrained in the WHSCM, it nevertheless behaves in a qualitatively similar manner in the real and null model networks.

However, both the unweighted and weighted clustering coefficients are much lower in the null models than in the real networks. To quantify this difference a bit further, we average the unweighted and weighted clustering coefficients across all nodes in the real networks and across all nodes in all the 10 WHSCM replicas of the real networks. The results are shown in Table~\ref{tab:real-nets-clustering}.
\begin{table}
\caption{The unweighted and weighted average local clustering coefficients $\boldsymbol{\bar{c}_{u,r}}$, $\boldsymbol{\bar{c}_{w,r}}$ in the the considered real weighted networks and in their synthetic WHSCM counterparts, $\boldsymbol{\bar{c}_{u,s}}$, $\boldsymbol{\bar{c}_{w,s}}$
}\label{tab:real-nets-clustering}
\begin{tabular}{lllll}
\hhline{=====}
\textbf{Network name} & $\boldsymbol{\bar{c}_{u,r}}$ & $\boldsymbol{\bar{c}_{u,s}}$ & $\boldsymbol{\bar{c}_{w,r}}$ & $\boldsymbol{\bar{c}_{w,s}}$ \\ 
\hline        
\rule{0pt}{3ex}\textit{C. elegans} metabolic  & $6.5 \cdot 10^{-1}$ & $2.2 \cdot 10^{-2}$ & $6.6 \cdot 10^{-3}$ & $2.8 \cdot 10^{-3}$ \\
Comp.\ geo.\ collab.           & $4.9 \cdot 10^{-1}$ & $3.2 \cdot 10^{-2}$ & $4.7 \cdot 10^{-3}$ & $8.2 \cdot 10^{-5}$ \\
Bible proper nouns             & $7.2 \cdot 10^{-1}$ & $4.6 \cdot 10^{-2}$ & $5.2 \cdot 10^{-3}$ & $5.0 \cdot 10^{-4}$ \\
\hhline{=====}
\end{tabular}
\end{table}
We see that in all the three real networks, both the unweighted and weighted average clustering is much higher, often by orders of magnitude, than in their null model counterparts. This observation suggests that the clustering coefficient is a statistically significant structural feature of these networks that cannot be explained by power-law degree and strength distributions alone. Some other mechanisms, such as latent geometry~\cite{allard2017geometric}, are thus responsible for the formation of clustering in these real networks.

\section{Discussion}\label{sec:discussion}

There exist a plenty of weighted network models with tunable degree and strength distributions~\cite{yook2001weighted,barrat2004weighted,barrat2004modeling,bianconi2005emergence,antal2005weight,wang2005mutual,ou2007power,tanaka2008weighted,krings2009urban,fagiolo2010international,barthelemy2011spatial,simini2012universal,popovic2012geometric,allard2017geometric}. Here, we have contributed to this list by introducing the unique unbiased null model, the WHSCM, that satisfies the maximum entropy requirement. The model produces random graphs whose entropy is maximal across all graphs whose joint distribution of degrees and strengths converges to a given distribution. The outline of the proof that this is indeed so in Appendix~\ref{sec:proof} is not as detailed as the proof of the maximum entropy properties of the HSCM in Ref.~\cite{van2018sparse}. The WHSCM proof can thus be improved by filling in all the missing details.

In developing a particular version of the model with power-law degree and strength distributions, we encountered the major challenge in the form of the system of nonlinear integral equations~(\ref{eq:whscm-bar(k)}),~(\ref{eq:whscm-bar(s)}) or (\ref{eq:whscm-exp-degree})~,(\ref{eq:whscm-exp-strength}) that need to be solved to find the latent parameter distribution $\rho(\nu,\mu)$ or $\rho(\lambda,\mu)$. These equations appear intractable, so that we devised a workaround that worked well. Yet one may definitely question how good our approach is in general, and how valid, reliable, and accurate the double power-law modeling of the distributions in Fig.~\ref{fig:mle-figure} is in particular. We dedicated Appendix~\ref{sec:app-scalings} to support our double power-law assumption, but the argument there is not very rigorous.
Is there a better, more accurate model for these distributions, possibly improving the convergence speeds in Fig.~\ref{fig:finite-n-exponent-estimates}? More generally, is there an entirely different, more principled approach to the problem of finding (approximate) solutions of the systems of equations (\ref{eq:whscm-bar(k)})~,(\ref{eq:whscm-bar(s)}), (\ref{eq:whscm-exp-degree})~,(\ref{eq:whscm-exp-strength})?

It would be nice to have such an approach because one may wish to constrain the degree-strength distributions not necessarily to power laws but to something else---to truncated power laws, for instance, observed in real weighted networks~\cite{newman2001structure,barrat2004architecture,da2009structural,zhang2010evolution}. What is the latent parameter distribution in this case?
In Fig.~\ref{fig:simple-pl-whscm}, we saw that a ``clean'' power-law choice of the latent parameter distribution $\rho(\lambda,\mu)$ led to truncated power laws for the marginals of $P(k,s)$, but even with this clean choice of $\rho(\lambda,\mu)$, things are difficult to control analytically as Appendix~\ref{sec:app-single-exponent-whscm} shows. In any case, for any desired strength-degree distribution $\rho(\kappa,\sigma)$, the most principled solution is the exact solution of the system of integral equations (\ref{eq:whscm-exp-degree}) and (\ref{eq:whscm-exp-strength}) with respect to the latent parameter distribution~$\rho(\lambda,\mu)$.

We emphasize that the introduced null model is what it is, a null model, so that it should be used as such. It should not be confused with or considered as a realistic model of real-world weighted networks. We saw, for instance, that the model does not capture clustering observed in real networks. Latent geometry was proposed in Ref.~\cite{allard2017geometric} as a possible mechanism explaining clustering in real weighted networks. It would be interesting to see whether the model in Ref.~\cite{allard2017geometric} satisfies the maximum entropy requirements, and if so, then under what constraints.

Related to that, it would be also nice to have a weighted generalization of random hyperbolic graphs~\cite{krioukov2010hyperbolic} whose maximum entropy properties are well understood and whose infinite temperature limit is exactly the HSCM~\cite{boguna2020small,aldecoa2015hyperbolic}. In other words, what is the model of weighted random hyperbolic graphs that have analogous maximum entropy properties and whose infinite temperature limit is the WHSCM?

\begin{acknowledgments}
We thank G.~Cimini, D.~Garlaschelli, \mbox{R.~Mastrandrea}, A.~Allard, M. {\'A}.~Serrano, and M.~Bogu{\~n}{\'a} for useful discussions and suggestions. This work was supported by NSF Grant No.~IIS-1741355 and ARO Grant Nos.~W911NF-17-1-0491 and W911NF-16-1-0391. M.~Kitsak was additionally supported by the NExTWORKx project. \mbox{F.~Papadopoulos} acknowledges support by the \mbox{TV-HGGs} project \mbox{(OPPORTUNITY/0916/ERC-CoG/0003)}, funded by the Cyprus Research and Innovation Foundation.
\end{acknowledgments}

\onecolumngrid
\newpage
\twocolumngrid

\appendix

\section{(W)HSCM as entropy maximizers}\label{sec:proof}

Here we summarize the key points of the proof from Ref.~\cite{van2018sparse} that the HSCM random graphs maximize graph entropy across all graphs whose degree distribution converges to a given distribution, and show how to generalize this proof to weighted graphs in the WHSCM.

\subsection{HSCM as entropy maximizer}

For a given distribution $P(k)$ of node degrees $k$, what is the graph ensemble whose ensemble distribution $P(G)$ of graphs $G$ maximizes Shannon entropy~(\ref{eq:shannon-entropy}) but in such a way that the degree distribution in the ensemble converges to $P(k)$? As proved in Ref.~\cite{van2018sparse}, the unique answer is the HSCM.

The proof is not exactly trivial because any brute-force attack at it is doomed to fail since if understood literally, the task is an intractable combinatorial optimization problem, a mission impossible. To circumvent this impasse, a workaround collection of techniques, based on the graphon theory~\cite{janson2013graphons}, was devised in Ref.~\cite{van2018sparse}. We describe this collection here.

The main idea of this workaround proof is to maximize not the entropy $S[P(G)]$ of the intractable ensemble distribution $P(G)$ but the graphon entropy defined below. Intuitively, the graphon entropy is the entropy of random edges conditioned on given values of the Lagrange multipliers $\nu_i$. In other words, for a given collection of fixed $\nu_i$s, the graphon entropy is the SCM entropy. The maximization of the graphon entropy turns out to be a tractable functional analysis problem with an explicit unique solution, but there is another contribution to the HSCM entropy which is the entropy of $\nu_i$s that are not fixed but random in the HSCM. The crux of the proof is to show that the entropy of the graphon that maximizes the graphon entropy is the leading term in the graph entropy $S[P(G)]$, while the entropy of $\nu_i$s is subleading. Since so, the graphon that maximizes the graphon entropy maximizes also the graph entropy, thus solving the original entropy maximization problem. We provide some key details behind how this works next.

In the HSCM, the graphon is no mystery but just the connection probability function $p(\nu,\nu')$. This function literally says that if the Lagrange multipliers of two nodes happen to be $\nu$ and $\nu'$, then the link between them is the Bernoulli random variable with the success rate~$p(\nu,\nu')$. Observe that the Bernoulli random variable $\mathrm{Be}(p)$ is trivially the random variable that maximizes the Shannon entropy of distributions on $\{0,1\}$ with mean $p$. Recall that the Lagrange multipliers $\nu$ as random variables can be mapped to the expected degrees $\kappa$ via (\ref{eq:hscm-kappa(nu)}) and (\ref{eq:hscm-kappa-abuse}) resulting in the graphon $p(\kappa,\kappa')$ expressed as a function of $\kappa$s. Observe that if $\kappa$ is now treated as a latent variable, then the expected degree of a node with latent variable $\kappa$ is $\kappa$:
\begin{equation}\label{eq:appA-kappa}
  \kappa = n \int p(\kappa,\kappa')\rho(\kappa')\,d\kappa'.
\end{equation}

The edges $a_{ij}$ in the HSCM are Bernoullis with different success rates $p(\kappa_i,\kappa_j)$ that are random because $\kappa$s are random. The entropy $S[\mathrm{Be}(p)]$ of the Bernoulli random variable with the success rate $p$ is
\begin{equation}
  S[\mathrm{Be}(p)]=-p\log{p}-(1-p)\log(1-p).
\end{equation}
These observations justify the definition of the entropy $S[p]$ of the graphon $p()$ in Ref.~\cite{janson2013graphons} which is
\begin{equation}\label{eq:appA-graphon-entropy}
  S[p]=\iint S[\mathrm{Be}(p(\kappa,\kappa'))] \rho(\kappa) \rho(\kappa^\prime) \,d\kappa\,d\kappa'.
\end{equation} 

What is the graphon $p()$ that maximizes the graphon entropy in Eq.~\eqref{eq:appA-graphon-entropy} while satisfying the constraint in Eq.~\eqref{eq:appA-kappa}, where $\rho(\kappa)$ is our desired expected degree distribution, the one that yields the desired $P(k)$? As shown in Ref.~\cite{van2018sparse}, it is relatively simple to prove that the unique exact answer is the Fermi-Dirac graphon in Eq.~\eqref{eq:scm-conn-prob} with $\nu$s mapped to $\kappa$s via (\ref{eq:hscm-kappa(nu)}) and (\ref{eq:hscm-kappa-abuse}). As also shown in Ref.~\cite{van2018sparse}, in sparse graphs this exact solution is asymptotically equivalent to the approximate expressions in Eqs.~(\ref{eq:scm-conn-prob-kappa}) and (\ref{eq:scm-conn-prob-cl}) that express the solution graphon explicitly in terms of the $\kappa$ variables.

Proving that the entropy of this graphon $S[p]$ is the dominating term in the graph entropy $S[P(G)]$, in comparison to entropy of random $\kappa$s, is a much more delicate endeavor. For this, the following techniques from~\cite{janson2013graphons} are properly adjusted in Ref.~\cite{van2018sparse}.

First, it is easy to see that the graphon entropy is a trivial lower bound for the HSCM graph entropy divided by $n\choose2$.
Indeed, if all $\kappa$s are fixed, then the graphon entropy in the HSCM is the entropy of the SCM graphs with this graphon divided by $n\choose2$.
The hard part is thus to find a matching upper bound, and this is where the techniques from Ref.~\cite{janson2013graphons} come really useful.

The key point in establishing such an upper bound is to recognize that for any partition of the values of $\kappa$s into consecutive intervals $\pi_k$, $k=1,\ldots,K$, the entropy of $P(G)$ is upper-bounded by the entropy of the averaged graphon defined below, plus the entropy of the indicator random variables $I_{ik}$ that indicate whether the random expected degree $\kappa_i$ of node $i$ happened to land in the interval $\pi_k$. The averaged graphon is defined as the piecewise constant function $\bar{p}(\kappa,\kappa')$ whose values for the values of $\kappa,\kappa'$ belonging to a given rectangle $\pi_k\times\pi_{k'}$ in the $\kappa\times\kappa'$ partition are equal to the average value of the graphon $p(\kappa,\kappa')$ in this rectangle.
Observe that the smaller the number of the partition intervals $K$, the smaller the
total entropy of the indicator random variables $I_{ik}$, just because there are fewer of them, but the larger the sum of the error terms coming
from graphon averaging, simply because rectangles $\pi_k\times\pi_{k'}$ are large. The smaller they are, the smaller
the total graphon entropy error term, but the larger the total entropy of indicators $I_{ik}$. The crux of the proof is to
find a ``sweet spot''---the right number of intervals of the right size guaranteeing the proper balance between
these two types of contributions to the upper bound, which we want to be tighter than the
difference between the graph and graphon entropies.
In sparse graphs, this task turns out to be a blade runner exercise.

Notwithstanding these blade runner difficulties, the required partition $\pi_k$ was found in~\cite{van2018sparse}, completing the proof that the HSCM is indeed the unique entropy maximizer across all random graph ensembles whose degree distribution converges to a given $P(k)$.

\subsection{WHSCM as entropy maximizer}

The maximum entropy HSCM proof described in the previous section should apply to the WHSCM as well, upon the modifications that we discuss below. The key idea behind these modifications is a proper generalization of graphons and their entropy to weighted networks.

Similar to the unweighted case, in the weighted case it is more convenient to deal with the expected degree and strength variables $\kappa,\sigma$ instead of the Lagrange multipliers $\nu,\mu$. The map from the latter to the former is given by Eqs.~(\ref{eq:whscm-bar(k)}) and (\ref{eq:whscm-bar(s)}) which, if rewritten in the $\kappa,\sigma$ variables, become the system of the self-consistency equations
\begin{align}
    \kappa &= n \iint p(\kappa,\sigma,\kappa',\sigma') \rho(\kappa',\sigma') \, d\kappa' \, d\sigma',\label{eq:appA-bar(k)}\\
    \sigma &= n \iint \omega(\kappa,\sigma,\kappa',\sigma') \rho(\kappa',\sigma') \, d\kappa' \, d\sigma',\label{eq:appA-bar(s)}
\end{align}
analogous to Eq.~\eqref{eq:appA-kappa}.

In unweighted networks, graph edges $a_{ij}$ are Bernoullis with different random success rates $p(\kappa_i,\kappa_j)$:
\begin{equation}
  a_{ij}=\mathrm{Be}[p(\kappa_i,\kappa_j)].
\end{equation}
In weighted networks, the edges are no longer Bernoullis. Instead they are random variables that we call Bernoulli Exponential, or BeExp for short:
\begin{equation}\label{eq:appA-wij}
  w_{ij}=\mathrm{BeExp}[p(\kappa_i,\sigma_i,\kappa_j,\sigma_j),\omega(\kappa_i,\sigma_i,\kappa_j,\sigma_j)],
\end{equation}
where $p$ and $\omega$ are the two parameters of the BeExp.

We define the vanilla BeExp as follows: if $w=\mathrm{BeExp}(p,\omega)$, where $p\in[0,1]$, $\omega>0$, and $w\geq0$, then $w=0$ with probability $1-p$, while with probability~$p$, $w$ is the exponential random variable with mean~$\omega$ (or rate $1/\omega$). In other words, the PDF of the BeExp $w=\mathrm{BeExp}(p,\omega)$ is
\begin{equation}
  P(w)=
  \begin{cases}
    1-p, & \mbox{if } w=0, \\
    \frac{p}{\omega}e^{-w/\omega}, & \mbox{if } w>0,
  \end{cases}
\end{equation}
so that its entropy is
\begin{align}
  S[\mathrm{BeExp}(p,\omega)] &=-\int_{0}^{\infty} P(w)\log{P(w)}\,dw\nonumber \\
  &= -p\log{p}-(1-p)\log{(1-p)}\nonumber\\
  &\quad +p(1+\log{\omega}) \nonumber\\
  &= S[\mathrm{Be}(p)]+pS[\mathrm{Exp}(\omega)],
\end{align}
where $S[\mathrm{Exp}(\omega)]=1+\log{\omega}$ is the entropy of the exponential distribution with mean $\omega$.

We observe that the $\mathrm{BeExp}(p,\omega)$ can be intuitively thought of as a ``smearing'' of the probability $p$ of $1$ (edge existence) in $\mathrm{Be}(p)$ into the $\mathrm{Exp}(\omega)$, the exponential distribution with mean $\omega$ (edge weight).
As $\mathrm{Be}(p)$ is trivially the maximum entropy distribution on $\{0,1\}$ with mean $p$, so is $\mathrm{BeExp}(p,\omega)$, less trivially, the maximum entropy distribution on $[0,\infty)$ with mean $\omega$ and $P(w>0)=p$. That is, the $\mathrm{BeExp}(p,\omega)$ is the maximum entropy distribution under the constraints that the edge exists with probability $p$ and that its mean weight is $\omega$. It is common knowledge in statistical mechanics that in maximum entropy canonical ensembles of systems of particles, the distributions of particles over particle states are also maximum entropy (Fermi-Dirac or Bose-Einstein). Since graph edges are analogous to particles in statistical mechanics~\cite{park2004statistical,garlaschelli2009generalized,gabrielli2019grand,cimini2019statistical}, these observations motivate us to constrain the space of all possible probability distributions on $[0,\infty)$ to the two-parametric maximum entropy BeExp family.

Equation~\eqref{eq:appA-wij} says that all edges in our weighted networks are BeExp's, albeit with different random parameters which are functions $p(\kappa,\sigma,\kappa',\sigma')$ and $\omega(\kappa,\sigma,\kappa',\sigma')$ of random $\kappa,\sigma$. Similar to the unweighted case, these observations instruct us to define the weighted graphon to be the parameters of our maximum entropy BeExp random variable. That is, we define a weighted graphon as \emph{the pair of functions} $\{p(\kappa,\sigma,\kappa',\sigma'), \omega(\kappa,\sigma,\kappa',\sigma')\}$. Similar to the unweighted graphon entropy in Eq.~\eqref{eq:appA-graphon-entropy}, we then define the weighted graphon entropy as
\begin{equation}\label{eq:appA-graphon-entropy-weighted}
\begin{split}
   S[p,\omega] &= \iiiint S\Bigl\{\mathrm{BeExp}\Bigl[p(\kappa,\sigma,\kappa',\sigma'),\omega(\kappa,\sigma,\kappa',\sigma')\Bigr]\Bigr\}\\
     &\times \rho(\kappa,\sigma)\,\rho(\kappa',\sigma')\,d\kappa\,d\sigma\,d\kappa'\,d\sigma'.
\end{split}
\end{equation}

The rest of the proof then proceeds as in the unweighted case, using the same techniques as in Ref.~\cite{van2018sparse}: first show that the unique graphon maximizing the graphon entropy in Eq.~\eqref{eq:appA-graphon-entropy-weighted} subject to the constraints in Eqs.~(\ref{eq:appA-bar(k)}) and (\ref{eq:appA-bar(s)}) is given by Eqs.~(\ref{eq:wscm-conn-prob}) and (\ref{eq:whscm-omega}) with $\nu,\mu$ mapped to $\kappa,\sigma$ via Eqs.~(\ref{eq:whscm-bar(k)}) and (\ref{eq:whscm-bar(s)}), and then prove that the entropy of this graphon dominates the graph entropy, while the entropy of random $\kappa,\sigma$ is negligible.
The latter step could be challenging as it calls for repeating the blade runner partition finding exercise from Ref.~\cite{van2018sparse}, this time for the product space of $\kappa \times \sigma$ values. This should be still possible using the same ideas as in Ref.~\cite{van2018sparse}---roughly, the key idea is that the partition is such that all its boxes have the same number of nodes in them on average. However, for the specific power-law WHSCM considered in this paper, or for any other WHSCM version in which strengths $\sigma$ are set to be a deterministic function of degrees $\kappa$, we only need to partition the space of $\kappa$ values. That is, the settings are exactly as in Ref.~\cite{van2018sparse} in that regard, so that exactly the same partition as in Ref.~\cite{van2018sparse} can be used in these cases.

\section{Simulation results for the relation between expected and actual degrees and strengths}\label{sec:app-concentrations}

\begin{figure}[htp]
 \centering
 \includegraphics[width=.49\textwidth]{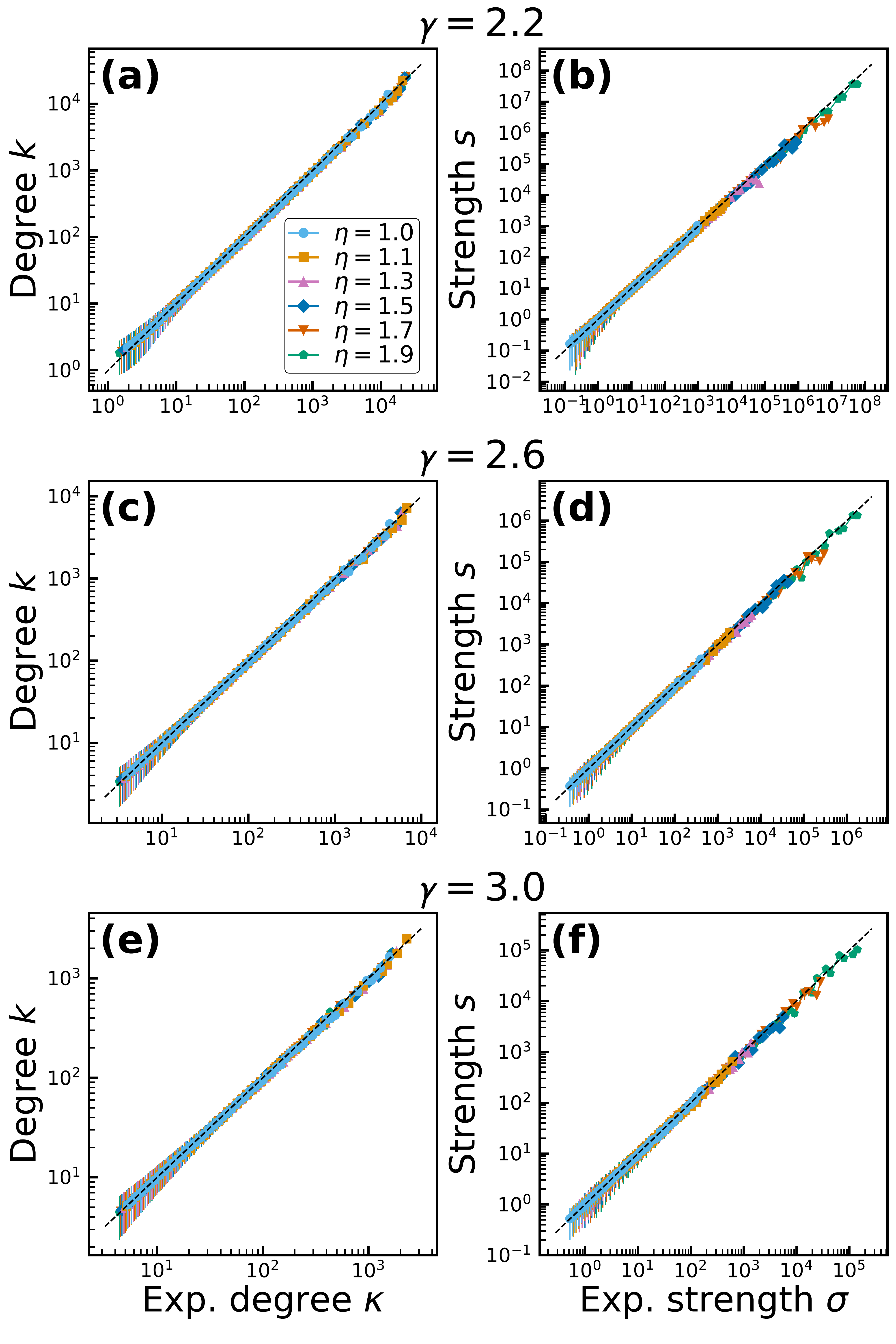}
  \caption{
  {\bf Correlation of the expected and actual degrees/strengths in the WHSCM.}
  Each row corresponds to the WHSCM input parameter $\gamma$ indicated on the top. Each color-coded plot corresponds to an $\eta$ parameter indicated in the legend. The rest of the WHSCM input parameters are $n=10,000$, $\bar{k} = 10$, and $\sigma_0 = 0.1$. Data is log-binned. Error bars show standard deviation from the bin mean. Black dashed line shows perfect linear correlation.
  }
  \label{fig:exp-vs-real}
\end{figure}

The WHSCM is formulated in terms of constraints for the joint distribution of \textit{expected} degrees and strengths, while in many cases we are interested in the behavior of \textit{actual} degrees and strengths realized in the corresponding ensemble of graphs. Thus, we need to show that the results obtained so far for expected values also hold for actual degrees and strengths. For latent variable graph models, it is known that actual degrees are concentrated around their expected values, and distributed according to Poisson distribution, i.e., $P(k|\kappa) = \mathrm{Pois}(\kappa)$~\cite{boguna2003class}. Since no similar claims are known for the behavior of strengths, we test the correlation between both $\kappa$ and $k$, and $\sigma$ and $s$ values in our model. To this end, we constructed log-binned scatter plots of actual degrees/strengths versus their expected values in WHSCM graphs with varying parameters $\gamma$ and $\eta$. The results are shown in Fig.~\ref{fig:exp-vs-real}. From the figure, it is evident that both degrees and strengths are highly correlated with their expected values. The narrow error bars also indicate that the distribution of actual degree/strength values around their expected values are narrow. Thus, the power-law scalings obtained for the expected values should also hold for actual values, as we have already demonstrated in the main text for various values of $\gamma$ and $\eta$.

\section{Scaling of $R$ with the number of nodes}\label{sec:app-R-a-scaling}

In the unweighted HSCM, the parameter $R$ scales as $R \sim \frac{1}{2} \log{n}$ with the network size $n$, which is evident from Eq.~\eqref{eq:hscm-kappa(nu)-approx}.
This motivated us to assume a similar scaling for the analysis of the WHSCM. In this appendix, we validate this choice by studying numerically how the parameters $R$ and $a$ scale with the system size in the WHSCM.

\begin{figure}[htp]
 \centering
 \includegraphics[width=.49\textwidth]{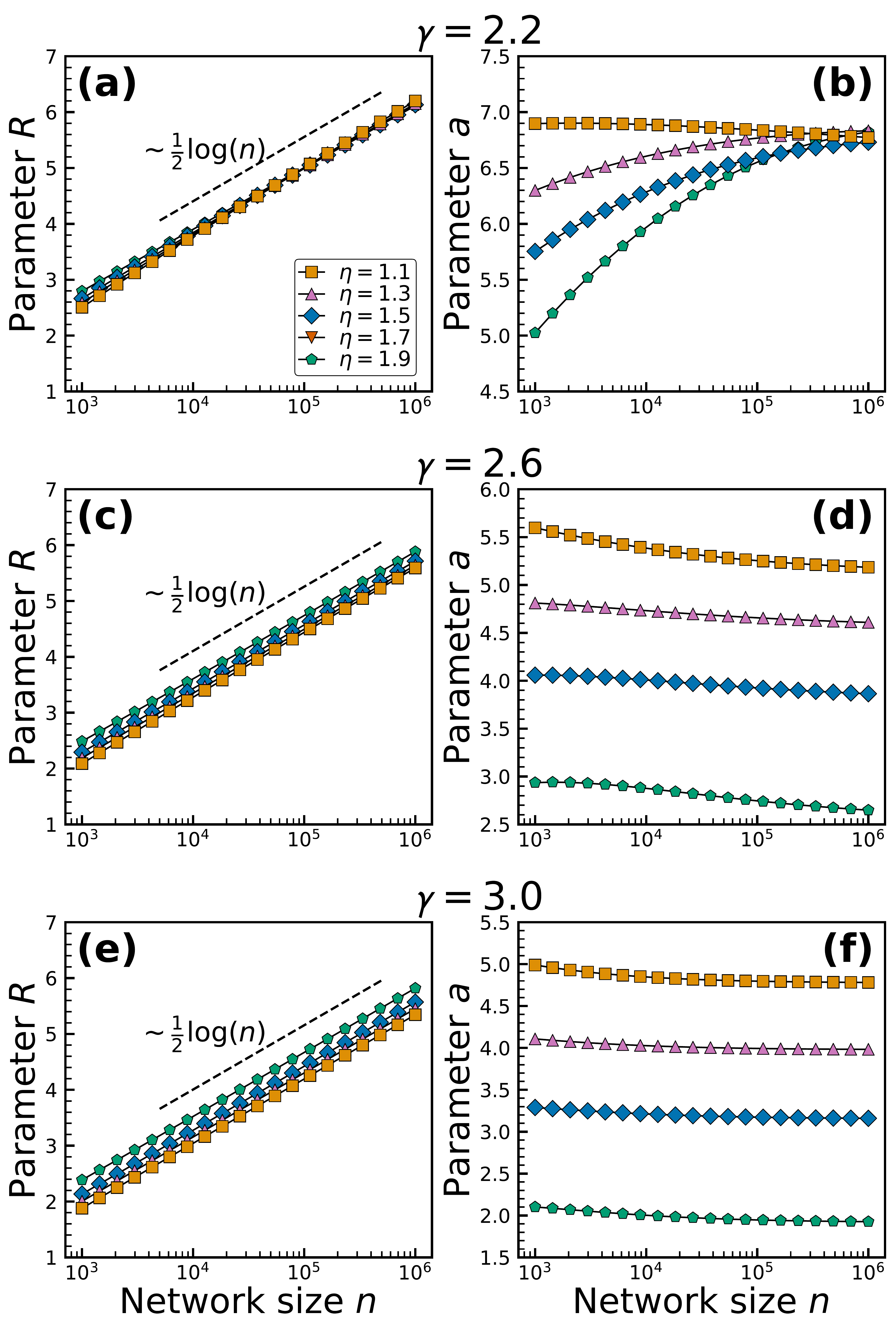}
  \caption{
  {\bf Scaling of the $R$ and $a$ parameters as a function of network size $n$ for various values of $\gamma$, $\eta$, and fixed values of $\bar{k} = 10$, and $\sigma_0 = 0.1$.}
  }
  \label{fig:R-a-solutions}
\end{figure}

To this end, we solve Eqs.~\eqref{eq:whscm-average-k} and~\eqref{eq:whscm-sigma0} as function of network size $n$ for various input exponents $\gamma$, $\eta$, and fixed values of average degree $\bar{k} = 10$, and expected strength-degree scaling constant $\sigma_0 = 0.1$. The resulting solution curves are shown in Fig.~\ref{fig:R-a-solutions}. The solutions indicate that the parameter $R$ scales with $n$ in the same way as in the unweighted HSCM case, while the parameter $a$ varies slowly with the network size. We note that since we solve for the $R$, $a$ numerically with fixed average degree requirement, the resulting networks are guaranteed to have constant average degree independent of network size, thus forming a sparse ensemble of graphs.

\section{WHSCM with power-law $\rho(\lambda, \mu)$}\label{sec:app-single-exponent-whscm}

Here we show that the WHSCM with the pure power-law joint distribution of latent parameters
\begin{align}
    \rho(\lambda, \mu) &= \rho(\lambda) \delta(\mu - f(\lambda)),\\
    \rho(\lambda) &= (\alpha - 1) \lambda^{-\alpha},\,\alpha > 2,\\
    f(\lambda) &= a \lambda^{-\beta},\,\beta \geq 0
\end{align}
does \textit{not} produce weighted networks with clean power-law distributions of degrees and strengths. We show this for the degree distribution $P(k)$. Similar arguments apply to the strength distribution $P(s)$.  

The expected degree of a node with the latent variable $\lambda$ is given by
\begin{equation}
    \kappa(\lambda) = n \int_{1}^{\infty} \frac{(\alpha - 1) \lambda^{\prime -\alpha}}{1 + a e^{2R} \Bigl[ \frac{\lambda^{-\beta} + \lambda^{\prime -\beta}}{\lambda \lambda^{\prime}} \Bigr]} d\lambda^{\prime}.
\end{equation}
While this integral is not expressible in a closed form in general, it is possible to find its approximations for different values of parameters $\lambda$ and $\beta$. For large $\beta$ and $\lambda \rightarrow 1$, the approximation is
\begin{equation}
    \begin{split}
    n& \int_{1}^{\infty} \frac{(\alpha - 1) \lambda^{\prime -\alpha}}{a e^{2R} \Bigl[ \frac{\lambda^{-\beta} + \lambda^{\prime -\beta}}{\lambda \lambda^{\prime}} \Bigr]} d\lambda^{\prime} = \frac{n}{a e^{2R}} \left( \frac{\alpha-1}{\alpha-2} \right) \lambda^{1+\beta} \cdot\\
    \cdot& {_2}F_1 \left(1, \frac{\alpha-2}{\beta}, 1 + \frac{\alpha-2}{\beta}, -\lambda^{\beta} \right).
    \end{split}
\end{equation}
For large $\lambda$, the approximation is
\begin{equation}
    \begin{split}
    n& \int_{1}^{\infty} \frac{n (\alpha-1) \lambda^{\prime -\alpha}}{1 + \frac{a e^{2R}}{\lambda \lambda^{\prime 1 + \beta}}} d\lambda^{\prime} = \\
    =n& \, {_2}F_1 \left( 1, \frac{\alpha - 1}{1 + \beta}, \frac{\alpha + \beta}{1 + \beta}, -\frac{a e^{2R}}{\lambda} \right).
    \end{split}
\end{equation}
These two approximations have different scalings of $\kappa(\lambda)$ with $\lambda$ that can be approximated by double power laws with two different exponents $\phi_1 \geq \phi_2 > 0$. While we are not able to obtain closed-form expressions for the exponents $\phi_1, \phi_2$ in terms of the model parameters $\alpha, \beta$ in general case, from numerical simulations we observe that networks generated according to this version of the WHSCM have degree distributions with double-power-law-like behavior. At least the degree distribution does definitely not look like a power but as a power law with a power-law cutoff as Fig.~\ref{fig:simple-pl-whscm} demonstrates.

In other words, a single power law distribution with exponent $\alpha$ of the latent parameter $\lambda$ results in two different scaling exponents for the resulting degree distribution $P(k)$, that are equal to $-\left(1 + (\alpha-1)/\phi_1\right)$ for small degrees $k$, and $-\left(1 + (\alpha-1)/\phi_2\right)$ for large degrees $k$. Such double-scaling behavior is yet another motivation to use double power law $\rho(\lambda)$ that would match the two scalings and result in single scaling of $P(k)$ with single target exponent $\gamma$.

\section{Analysis of expected degrees and strengths in the WHSCM}\label{sec:app-scalings}

In the WHSCM, each node is characterized by the two latent parameters $\lambda$ and $\mu$ that are distributed according to the joint probability distribution from Eq.~\eqref{eq:whscm-joint-prob}. In Sec.~\ref{sec:whscm}, we stated expressions for the exponent $\alpha_1, \alpha_2, \beta_1$ and $\beta_2$ for the densities of $\lambda$ and $\mu$. They are obtained from analysis of the integral expressions for both $\kappa$ and $\sigma$ in Eqs.~\eqref{eq:whscm-exp-degree} and~\eqref{eq:whscm-exp-strength}, respectively.

\textbf{Approximating the integral expressions for expected degrees.} 
By our double power-law choice for the marginal distributions for $\lambda$ and $\mu$, the expected degree of a node with latent parameter $\lambda$, Eq.~\eqref{eq:whscm-exp-degree}, may be written as the following four integrals, where each integral represents one of combinations for ``small'' ($\leq \lambda_c$) and ``large'' ($> \lambda_c$) parameters $\lambda$ and $\lambda^{\prime}$:
\begin{align}
    I_1 (\lambda) &= n \int\limits_{1}^{\lambda_c} \frac{A_1 \lambda^{\prime -\alpha_1}}{1 + a e^{2R} \left( \frac{\lambda^{-\beta_1} + \lambda^{\prime -\beta_1}}{\lambda \lambda^{\prime}} \right)} d\lambda^{\prime}, \label{eq:whscm-i1} \\
    I_2 (\lambda) &= n \int\limits_{\lambda_c}^{\infty} \frac{A_2 \lambda^{\prime -\alpha_2}}{1 + a e^{2R} \left( \frac{\lambda^{-\beta_1} + \lambda_c^{\beta_2 - \beta_1} \lambda^{\prime -\beta_2}}{\lambda \lambda^{\prime}} \right)} d\lambda^{\prime}, \label{eq:whscm-i2} \\
    I_3 (\lambda) &= n \int\limits_{1}^{\lambda_c} \frac{A_1 \lambda^{\prime -\alpha_1}}{1 + a e^{2R} \left( \frac{\lambda_c^{\beta_2 - \beta_1}\lambda^{-\beta_2} + \lambda^{\prime - \beta_1}}{\lambda \lambda^{\prime}} \right)} d\lambda^{\prime}, \label{eq:whscm-i3} \\
    I_4 (\lambda) &= n \int\limits_{\lambda_c}^{\infty} \frac{A_2 \lambda^{\prime -\alpha_2}}{1 + a \lambda_c^{\beta_2 - \beta_1} e^{2R} \left( \frac{\lambda^{-\beta_2} + \lambda^{\prime - \beta_2}}{\lambda \lambda^{\prime}} \right)} d\lambda^{\prime}. \label{eq:whscm-i4}
\end{align}
While these integrals cannot be computed directly in a closed-form, it is possible to approximate them. For~\eqref{eq:whscm-i1}, we use that for both $\lambda, \lambda^\prime < \lambda_{c}$ the $1$ in the denominator can be removed. This cannot be done for the other three integrals, since $\lambda^\prime > \lambda_{c}$ or $\lambda > \lambda_{c}$. Therefore, in the second integral~\eqref{eq:whscm-i2}, we use that
$\lambda_{c}^{\beta_2 - \beta_1} \lambda^{\prime -\beta_2} \leq \lambda_{c}^{-\beta_1}$
and hence this term is negligible with respect to $\lambda^{-\beta_1}$
when $1 \le \lambda \le \lambda_{c}$. A similar approach is applied to~\eqref{eq:whscm-i3}, with the role of $\lambda$ and $\lambda^\prime$ reversed. Finally, for integral~\eqref{eq:whscm-i4}, we use that for $\lambda, \lambda^\prime > \lambda_{c}$,
\[
    \lambda_{c}^{\beta_2 - \beta_1} \left(\frac{\lambda^{-\beta_2} + \lambda^{\prime -\beta_2}}{\lambda \lambda^\prime}\right) < 2 \lambda_{c}^{-(2 + \beta_1)},
\]
and, given that $\lambda_c \approx (2 a e^{2R})^{1/(2+\beta_1)}$ and $a e^{2R} \sim n$, as we demonstrate in Appendix~\ref{sec:app-R-a-scaling}, we have
\[
    2 \lambda_{c}^{-(2 + \beta_1)} \approx \frac{1}{a e^{2R}} \ll 1,
\]
so that the $1$ is the dominant term in the denominator.

This allows us to obtain the following approximations for the integrals above:
\begin{align}
    I_1 (\lambda) &\approx n \int\limits_{1}^{\lambda_c} \frac{A_1 \lambda^{\prime -\alpha_1}}{a e^{2R} \left( \frac{\lambda^{-\beta_1} + \lambda^{\prime -\beta_1}}{\lambda \lambda^{\prime}} \right)} d\lambda^{\prime}, \\
    I_2 (\lambda) &\approx n \int\limits_{\lambda_c}^{\infty} \frac{A_2 \lambda^{\prime -\alpha_2}}{1 + a e^{2R} \left( \frac{\lambda^{-1-\beta_1}}{\lambda^{\prime}} \right)} d\lambda^{\prime}, \\
    I_3 (\lambda) &\approx n \int\limits_{1}^{\lambda_c} \frac{A_1 \lambda^{\prime -\alpha_1}}{1 + a e^{2R} \left( \frac{\lambda^{\prime -1 - \beta_1}}{\lambda} \right)} d\lambda^{\prime}, \\
    I_4 (\lambda) &\approx n \int\limits_{\lambda_c}^{\infty} A_2 \lambda^{\prime -\alpha_2} d\lambda^{\prime}.
\end{align}
The four integrals on the right-hand side can be evaluated exactly:
\begin{align}
    \begin{split}
    I_1 (\lambda) &\approx \frac{n A_1 \lambda^{1+\beta_1}}{a e^{2R} (\alpha_1 - 2)} \cdot \\
                  &\quad \cdot \biggl[ {_2}F_1 \left( 1, \frac{\alpha_1 - 2}{\beta_1}, 1 + \frac{\alpha_1 - 2}{\beta_1}, -\lambda^{\beta_1} \right) - \\
                  &\quad - \lambda_c^{2-\alpha_1} {_2}F_1 \left( 1, \frac{\alpha_1 - 2}{\beta_1}, 1 + \frac{\alpha_1 - 2}{\beta_1}, -\left( \frac{\lambda}{\lambda_c} \right)^{\beta_1} \right) \biggr], 
    \end{split}
    \label{eq:whscm-i1-approx} \\
    I_2 (\lambda) &\approx \frac{n A_2 \lambda_c^{1-\alpha_2}}{\alpha_2 - 1} {_2}F_1 \left( 1, \alpha_2 - 1, \alpha_2, -\frac{a e^{2R}}{\lambda_c \lambda^{1 + \beta_1}} \right) \label{eq:whscm-i2-approx} \\
    \begin{split}
    I_3 (\lambda) &\approx \frac{n A_1}{\alpha_1 - 1} \biggl[ {_2}F_1 \left( 1, \frac{\alpha_1 - 1}{1 + \beta_1}, \frac{\alpha_1 + \beta_1}{1 + \beta_1}, -\frac{a e^{2R}}{\lambda} \right) - \\
                  &\quad - \lambda_c^{1-\alpha_1} {_2}F_1 \left( 1, \frac{\alpha_1 - 1}{1 + \beta_1}, \frac{\alpha_1 + \beta_1}{1 + \beta_1}, -\frac{a e^{2R}}{\lambda_c^{1+\beta_1} \lambda} \right) \biggr],
    \end{split}
    \label{eq:whscm-i3-approx} \\
    I_4 (\lambda) &\approx \frac{n A_2 \lambda_c^{1-\alpha_2}}{\alpha_2 - 1}, \label{eq:whscm-i4-approx}
\end{align}
where ${_2}F_{1} (q_1, q_2, q_3, z)$ is Gauss hypergeometric function.

\textbf{Approximating the integral expressions for expected strengths.} 
As was the case for the expected degree, the expected strength as a function of the latent parameter $\lambda$, see Eq.~\eqref{eq:whscm-exp-strength}, may be split into four integrals as follows:
\begin{align}
    \begin{split}
    I_5 (\lambda) &= n \int\limits_{1}^{\lambda_c} \frac{A_1 \lambda^{\prime -\alpha_1}}{1 + a e^{2R} \left( \frac{\lambda^{-\beta_1} + \lambda^{\prime -\beta_1}}{\lambda \lambda^{\prime}} \right)} \cdot\\ 
    &\quad\quad\quad \cdot \frac{1}{a \left( \lambda^{-\beta_1} + \lambda^{\prime -\beta_1} \right)} d\lambda^{\prime}, 
    \end{split}
    \label{eq:whscm-i5} \\
    \begin{split}
    I_6 (\lambda) &= n \int\limits_{\lambda_c}^{\infty} \frac{A_2 \lambda^{\prime -\alpha_2}}{1 + a e^{2R} \left( \frac{\lambda^{-\beta_1} + \lambda_c^{\beta_2 - \beta_1} \lambda^{\prime -\beta_2}}{\lambda \lambda^{\prime}} \right)} \cdot \\
    &\quad\quad\quad \cdot \frac{1}{a \left( \lambda^{-\beta_1} + \lambda_c^{\beta_2 - \beta_1}\lambda^{\prime -\beta_2} \right)} d\lambda^{\prime},
    \end{split}
    \label{eq:whscm-i6} \\
    \begin{split}
    I_7 (\lambda) &= n \int\limits_{1}^{\lambda_c} \frac{A_1 \lambda^{\prime -\alpha_1}}{1 + a e^{2R} \left( \frac{\lambda_c^{\beta_2 - \beta_1}\lambda^{-\beta_2} + \lambda^{\prime - \beta_1}}{\lambda \lambda^{\prime}} \right)} \cdot \\
    &\quad\quad\quad \cdot \frac{1}{a \left( \lambda_c^{\beta_2-\beta_1} \lambda^{-\beta_2} + \lambda^{\prime -\beta_1} \right)} d\lambda^{\prime},
    \end{split}
    \label{eq:whscm-i7} \\
    \begin{split}
    I_8 (\lambda) &= n \int\limits_{\lambda_c}^{\infty} \frac{A_2 \lambda^{\prime -\alpha_2}}{1 + a \lambda_c^{\beta_2 - \beta_1} e^{2R} \left( \frac{\lambda^{-\beta_2} + \lambda^{\prime - \beta_2}}{\lambda \lambda^{\prime}} \right)} \cdot \\
    &\quad\quad\quad \cdot \frac{1}{a \lambda_c^{\beta_2-\beta_1} \left( \lambda^{-\beta_2} + \lambda^{\prime -\beta_2} \right)} d\lambda^{\prime}. 
    \end{split}
    \label{eq:whscm-i8}
\end{align}
Using arguments similar to those for the expected degree, we may find the following approximations for the integrals above:
\begin{align}
    \begin{split}
    I_5 (\lambda) &\approx \frac{n A_1 \lambda^{1 + \beta_1}}{e^{2R} a^2 \beta_1 \left(1 + \left( \lambda_c / \lambda \right)^{\beta_1} \right)} \biggl[ \frac{\lambda_c^{\beta_1} + \lambda^{\beta_1}}{1 + \lambda^{\beta_1}} - \\
                  &\quad - \lambda_c^{2 + \beta_1 - \alpha_1} + \frac{(2+\beta_1-\alpha_1) (\lambda_c^{\beta_1} + \lambda^{\beta_1})}{\alpha_1 - 2} \cdot \\
                  &\quad \cdot \biggl\{ {_2}F_1 \left( 1, \frac{\alpha_1 - 2}{\beta_1}, 1 + \frac{\alpha_1 - 2}{\beta_1}, -\lambda^{\beta_1} \right) - \\
                  &\quad - \lambda_{c}^{2-\alpha_1} {_2}F_{1} \left( 1, \frac{\alpha_1 - 2}{\beta_1}, 1 + \frac{\alpha_1 - 2}{\beta_1}, -\left(\frac{\lambda}{\lambda_{c}}\right)^{\beta_1} \right) \biggr\} \biggr],
    \end{split}
    \label{eq:whscm-i5-approx} \\
    I_6 (\lambda) &\approx \frac{n A_2 \lambda_c^{1-\alpha_2} \lambda^{\beta_1}}{a (\alpha_2 - 1)} {_2}F_{1} \left( 1, \alpha_2 - 1, \alpha_2, -\frac{a e^{2R}}{\lambda_c \lambda^{1+\beta_1}} \right), \label{eq:whscm-i6-approx} \\
    \begin{split}
    I_{7} (\lambda) &\approx \frac{n A_1}{a (\alpha_1 - \beta_1 - 1)} \cdot \\ 
    &\quad \cdot \biggl[ {_2}F_{1} \left( 1, \frac{\alpha_1}{1+\beta_1} - 1, \frac{\alpha_1}{1+\beta_1}, -\frac{a e^{2R}}{\lambda} \right) - \\
    &\quad - \lambda_c^{1+\beta_1-\alpha_1} {_2}F_{1} \left( 1, \frac{\alpha_1}{1+\beta_1} - 1, \frac{\alpha_1}{1+\beta_1}, -\frac{a e^{2R}}{\lambda_c^{1+\beta_1} \lambda} \right) \biggr],
    \end{split}
    \label{eq:whscm-i7-approx} \\
    \begin{split}
    I_{8} (\lambda) &\approx \frac{n A_2 \lambda_c^{1+\beta_1-\alpha_2-\beta_2} \lambda^{\beta_2}}{a (\alpha_2 - 1)} \cdot \\
    &\quad \cdot {_2}F_{1} \left( 1, \frac{\alpha_2 - 1}{\beta_2}, 1 + \frac{\alpha_2 - 1}{\beta_2}, -\left(\frac{\lambda}{\lambda_c}\right)^{\beta_2} \right).
    \end{split}
    \label{eq:whscm-i8-approx}
\end{align}

\begin{figure}[htp]
 \centering
 \includegraphics[width=.4\textwidth]{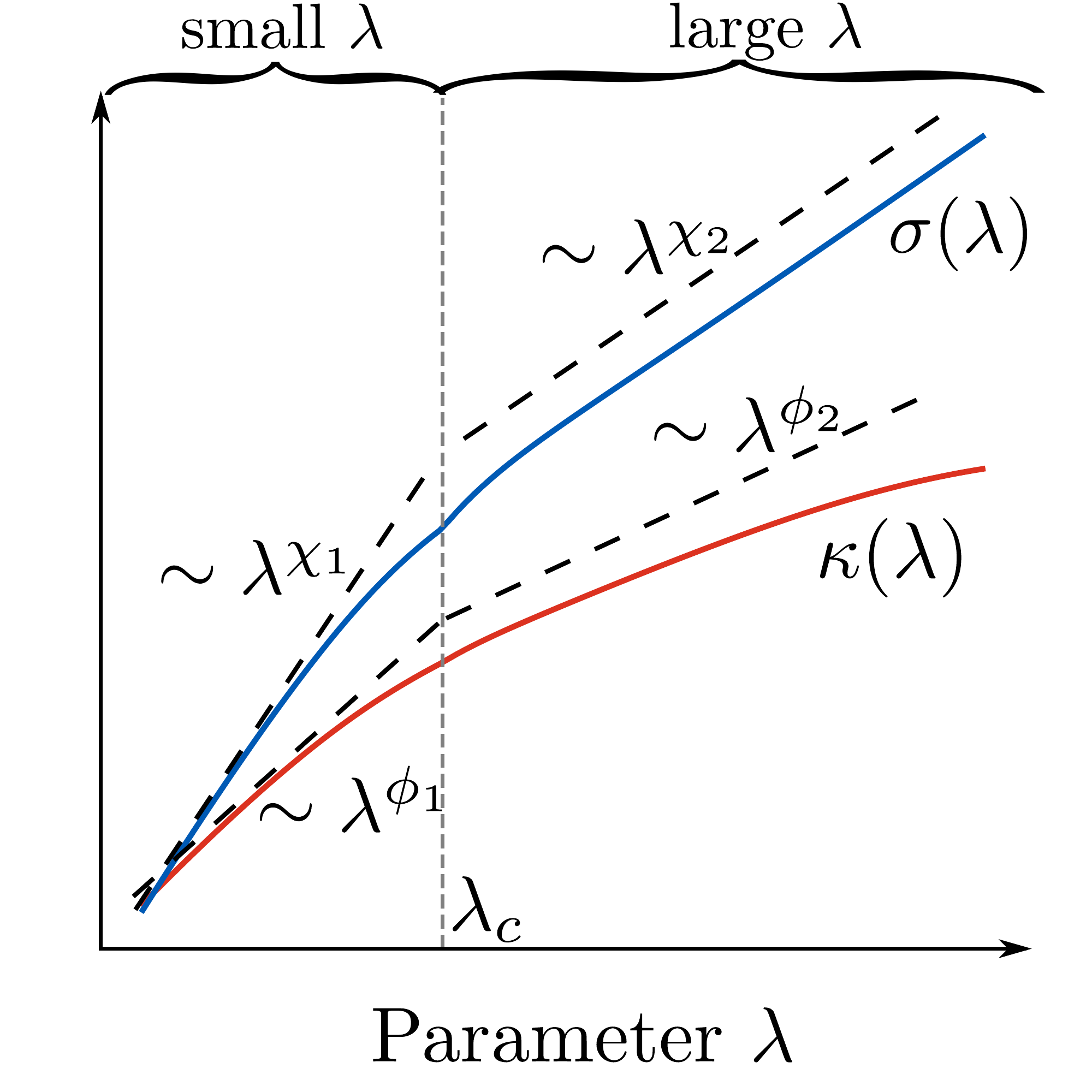}
  \caption{
  {\bf Illustration of the approximation used to find expression for $\kappa$ and $\sigma$.}
  The scheme is plotted on the log-log scale.
  }
  \label{fig:pl-approx-scheme}
\end{figure}

\textbf{Approximating the behavior of the integrals with power-law scalings.}
Unfortunately, it is hard to extract anything about the behavior of the $\kappa(\lambda)$ or $\sigma(\lambda)$ from the approximations that we obtained above. However, numerical evaluation of the expressions above, shows that the expected degree $\kappa (\lambda)$ scales roughly as some power of the latent parameter $\lambda$, where the scaling changes its exponent around the $\lambda_c$ point, as shown in Fig.~\ref{fig:pl-approx-scheme}. Similarly, the expected strength $\sigma(\lambda)$ scales roughly as some different power of the latent parameter $\lambda$, also changing the scaling exponent around the $\lambda_c$. We share the \textsc{mathematica} notebook \texttt{kappa-sigma-approximations.nb} to plot the approximated functions $\kappa(\lambda)$, $\sigma(\lambda)$ for various values of the model parameters at the GitHub repository~\cite{githubcode}.

We therefore seek for an approximation of the expected degree function in the double power-law form that changes its exponent at the constant $\lambda_c$, i.e.:
\begin{equation}\label{eq:whscm-kappa-scaling}
    \kappa (\lambda) \sim
    \begin{cases}
        \lambda^{\phi_1},\,\, 1 < \lambda \leq \lambda_c, \\
        \lambda^{\phi_2},\,\, \lambda_c < \lambda < \infty.
    \end{cases}
\end{equation}
Similarly, for the expected strength we have:
\begin{equation}\label{eq:whscm-sigma-scaling}
    \sigma (\lambda) \sim
    \begin{cases}
        \lambda^{\chi_1},\,\, 1 < \lambda \leq \lambda_c, \\
        \lambda^{\chi_2},\,\, \lambda_c < \lambda < \infty.
    \end{cases}
\end{equation}
With this ansatz, we now proceed to investigate how the scaling exponents $\phi_1, \phi_2, \chi_1, \chi_2$ behave as functions of the model parameters: $\alpha_1, \alpha_2, \beta_1, \beta_2, \gamma$ and $\eta$.

First, we note that the scaling exponents $\phi_1, \phi_2, \chi_1, \chi_2$ along with the model parameters $\alpha_1, \alpha_2, \beta_1, \beta_2$ should be related to the exponents $\gamma$ and $\delta = 1 + \frac{\gamma-1}{\eta}$ of the degree and strength power-law distributions. Indeed, given the distribution of the latent parameter $\lambda$ from Eq.~\eqref{eq:whscm-rho-lambda} and the $\kappa$ scaling from the Eq.~\eqref{eq:whscm-kappa-scaling}, we find that $\kappa$ is distributed according to $\rho(\kappa) \sim \kappa^{-(1 + (\alpha_1 - 1)/\phi_1)}$ for $\lambda \leq \lambda_c$, and $\rho(\kappa) \sim \kappa^{-(1 + (\alpha_2 - 1)/\phi_2)}$ for $\lambda > \lambda_c$. Similarly, using the Eqs.~\eqref{eq:whscm-rho-lambda} and~\eqref{eq:whscm-sigma-scaling}, the expected strength $\sigma$ is distributed according to $\rho(\sigma) \sim \sigma^{-(1 + (\alpha_1 - 1)/\chi_1)}$ for $\lambda \leq \lambda_c$, and $\rho(\sigma) \sim \sigma^{-(1 + (\alpha_2 - 1)/\chi_2)}$ for $\lambda > \lambda_c$. As we require that $\rho(\kappa) \sim \kappa^{-
\gamma}$ and $\rho(\sigma) \sim \sigma^{-(1 + (\gamma-1)/\eta)}$, the model parameters should satisfy:
\begin{align}
    \phi_1 &= (\alpha_1 - 1) / (\gamma - 1),\label{eq:exponents-phi1} \\
    \phi_2 &= (\alpha_2 - 1) / (\gamma - 1),\label{eq:exponents-phi2} \\
    \chi_1 &= \eta \phi_1,\label{eq:exponents-chi1} \\
    \chi_2 &= \eta \phi_2.\label{eq:exponents-chi2}
\end{align}
With these equations, we analyze how the scaling exponents $\phi_1, \phi_2, \chi_1, \chi_2$ depend on the parameters of the model.

\textbf{Analysis of the $\phi_1$ scaling exponent.} 
We start by analyzing the scaling of $\kappa \sim \lambda^{\phi_1}$ for the small values of $\lambda$, i.e., $\lambda \leq \lambda_c$. While it is possible to get the scaling of hypergeometric functions from Eqs.~\eqref{eq:whscm-i1-approx} and~\eqref{eq:whscm-i2-approx}, we note that the resulting scaling of $\kappa$ as a function of $\lambda$ is given by the sum of $\lambda$ terms raised to different powers. Moreover, coefficients in front of these terms may change their signs depending on the choice of the parameters $\alpha_1, \alpha_2, \beta_1, \beta_2$. In general, for \textit{any} combination of the model parameters, it is impossible to approximate this sum of power terms as a single power of $\lambda$, e.g., by extracting leading order terms. 

To circumvent this issue, we find an approximation to the $\phi_1$ scaling using the observations from the MLE inference of the WSCM latent parameters introduced in Sec.~\ref{sec:mle}. More precisely, we first infer the nodes' $\lambda$ and $\mu$ parameters given degree and strength sequences with predefined exponents $\gamma$ and $\eta$. Second, we obtain the estimate of the $\phi_1$ by linear fitting of the $\kappa(\lambda)$ function on the log-log scale. We observe that for various input values of $\gamma$ and $\eta$, the $\phi_1$ scaling exponent is very close to $\eta$. We therefore assume that $\phi_1 \approx \eta$, and, given Eq.~\eqref{eq:exponents-phi1}, the $\alpha_1$ parameter is set to:
\begin{equation}
    \alpha_1 \approx 1 + \eta (\gamma - 1). 
\end{equation}

\textbf{Analysis of the $\chi_1$ scaling exponent.}
The approximated expression for the expected strength $\sigma(\lambda)$ in the small regime, i.e., $\lambda \leq \lambda_c$, is given by Eqs.~\eqref{eq:whscm-i5-approx} and~\eqref{eq:whscm-i6-approx}. Although it is again possible to obtain $\sigma(\lambda)$ scaling in terms of series of various $\lambda$ power-terms, it is hard to extract a single power scaling that approximates the sum of these power-terms for all parameters $\alpha_1, \alpha_2, \beta_1, \beta_2$. However, it is possible to relate the $\chi_1$ scaling exponent to the $\beta_1$ parameter using the following considerations. From the MLE inference of the WSCM latent parameters as in the case of the $\phi_1$ scaling exponent, we observe that the resulting empirical model parameter $\beta_1$ is linearly related to the $\eta - 1$, and the coefficient between the two depends only on the input parameter $\gamma$, i.e., $\beta_1 \approx u(\gamma) (\eta - 1)$, where $u(\gamma)$ is a function of $\gamma$. By numerical fitting, we find that $u(\gamma)$ behaves approximately as $u(\gamma) \approx \left( \gamma - \frac{\gamma - 2}{\gamma} \right)$. Thus
\begin{equation}
    \beta_1 \approx \Bigl( \gamma - \frac{\gamma - 2}{\gamma} \Bigr) \Bigl(\eta - 1 \Bigr), 
\end{equation}
which yields $\beta_1 (\eta + 1) \approx ( \gamma - (\gamma - 2) / \gamma ) ( \eta^2 -1 )$, so that $\eta^2 \approx 1 + \frac{(\eta+1) \beta_1}{\gamma - (\gamma - 2)/ \gamma}$. Moreover, given Eq.~\eqref{eq:exponents-chi1} and the fact that $\phi_1$ may be approximated as $\phi_1 \approx \eta$, the resulting exponent $\chi_1$ should be approximately equal to $\chi_1 \approx \eta^2$. This means that $\chi_1 \approx 1 + \frac{(\eta+1) \beta_1}{\gamma - (\gamma - 2)/ \gamma}$.

\textbf{Analysis of the $\phi_2$ scaling exponent.} The scaling of $\kappa(\lambda) \sim \lambda^{\phi_2}$ for large $\lambda > \lambda_c$ is encoded in the two integrals from Eqs.~\eqref{eq:whscm-i3-approx} and~\eqref{eq:whscm-i4-approx}. The approximation in Eq.~\eqref{eq:whscm-i4-approx} does not have any $\lambda$-dependent terms, so it does not contribute to the scaling. We may approximate the scaling of the hypergeometric functions appearing in Eq.~\eqref{eq:whscm-i3-approx} for $\lambda \rightarrow \lambda_c$. In this case, the argument of the \textit{second} hypergeometric function approaches $-1$, therefore, the function approaches a constant and does not contribute to the scaling. The \textit{first} hypergeometric function may be approximated as follows, assuming that its argument is large:
\begin{equation}
    \begin{split}
    &{_2}F_1 \left( 1, \frac{\alpha_1 - 1}{1 + \beta_1}, 1 + \frac{\alpha_1 - 1}{1 + \beta_1}, -\frac{a e^{2R}}{\lambda} \right) \approx \\
    & \approx \frac{\alpha_1 - 1}{\alpha_1 - 2 - \beta_1} \left( \frac{\lambda}{a e^{2R}} \right) + \\
    & + \frac{\pi (\alpha_1 - 1)}{(1 + \beta_1) \sin{\left(\frac{\pi (\alpha_1 - 1)}{1 + \beta_1}\right)}} \left( \frac{\lambda}{a e^{2R}} \right)^{\frac{\alpha_1 - 1}{1+\beta_1}}.
    \end{split}
\end{equation}
Additionally, in the case when $\lambda \rightarrow \infty$, the $\kappa(\lambda)$ function should saturate to the maximum possible degree $n-1$, therefore, we only consider the $\kappa(\lambda)$ scaling near the left boundary of this regime, i.e., $\lambda_c$. The signs of the coefficients in front of the two $\lambda$-dependent terms may be of different signs, so it is again impossible to extract a single-exponent scaling from the expression above. However, from the WSCM MLE-inferred latent parameters as was done in the case of $\phi_1$, $\chi_1$ scalings, we observe that for large values of $\eta \gg 1$, the scaling exponent $\phi_2 \rightarrow \frac{\alpha_1 - 1}{1 + \beta_1}$. We therefore seek a scaling exponent $\phi_2$ in the form $\phi_2 \approx \psi(\gamma, \eta) \frac{\alpha_1 - 1}{1 + \beta_1}$. We know that $\phi_2 \rightarrow 1$ when $\eta \rightarrow 1$ to be compatible with the unweighted HSCM, and we know that in this limit $\beta_1 \rightarrow 0$, $\alpha_1 \rightarrow \gamma$. Therefore, there should be a prefactor of $\frac{1}{\gamma - 1}$ in front of the $\frac{\alpha_1 - 1}{1 + \beta_1}$ to guarantee that $\phi_2 \rightarrow 1$ in the HSCM limit. Additionally, as $\phi_2 \rightarrow \frac{\alpha_1 - 1}{1 + \beta_1}$ in the large $\eta$ limit, we need to compensate for this prefactor. Given these requirements, we find from the numerical fitting procedure that the following form of $\psi(\gamma, \eta)$ fits well the $\phi_2$ exponent for various $\gamma$, $\eta$ parameters: $\psi(\gamma, \eta) = \frac{1}{\gamma - 1} \bigl[ 1 + (\gamma - 2) \left(1 - 1/ \eta \right) \bigr]$. Then the $\phi_2$ scaling exponent is: $\phi_2 = \frac{\alpha_1 - 1}{1 + \beta_1} \frac{1}{\gamma - 1} \bigl[ 1 + (\gamma - 2) \left(1 - 1/ \eta \right) \bigr]$. Therefore, the model parameter $\alpha_2$ should be selected as follows:
\begin{equation}
    \alpha_2 \approx 1 + \frac{\alpha_1 - 1}{1 + \beta_1} \left( 1 + (\gamma - 2) (1 - 1 / \eta) \right).
\end{equation} 

\textbf{Analysis of the $\chi_2$ scaling exponent.} The scaling of $\sigma(\lambda)$ is given in Eq.~\eqref{eq:whscm-i7-approx} and~\eqref{eq:whscm-i8-approx}. The hypergeometric functions from Eq.~\eqref{eq:whscm-i7-approx} will not contribute to the scaling in the large $\lambda$ limit, as their arguments would approach $-1$, so we only expect to see an effect of these functions near the left boundary of the region, $\lambda_c$. Conversely, the hypergeometric function from Eq.~\eqref{eq:whscm-i8-approx} is the main contributor to the scaling for large $\lambda$. We note that we may not neglect the behavior of the $\sigma(\lambda)$ for large $\lambda$, as, in general, strengths are not bounded for a weighted network, unlike degrees that have to be at most $n-1$. Using the large argument approximation for hypergeometric functions as before, we obtain for Eq.~\eqref{eq:whscm-i8-approx}:
\begin{equation}
    \begin{split}
        & {_2}F_1 \left( 1, \frac{\alpha_2 - 1}{\beta_2}, 1 + \frac{\alpha_2 - 1}{\beta_2}, -\left( \frac{\lambda}{\lambda_c} \right)^{\beta_2} \right) \approx \\
        & \approx \frac{\alpha_2 - 1}{\alpha_2 - 1 - \beta_2} \left( \frac{\lambda}{\lambda_c} \right)^{-\beta_2} + \frac{\pi (\alpha_2 - 1)}{\beta_2 \sin{\left( \frac{\pi (\alpha_2 - 1)}{\beta_2} \right)}} \left( \frac{\lambda}{\lambda_c} \right)^{1 - \alpha_2}.
    \end{split}
\end{equation}
Given the $\lambda^{\beta_2}$ prefactor in Eq.~\eqref{eq:whscm-i8-approx}, we observe that the only possible resulting scaling may be $\sim \lambda^{1 + \beta_2 - \alpha_2}$. However, we note that the large-argument approximation used above is only valid for $\beta_1 > 0$ and $\beta_2 > 0$. Instead, when $\eta \rightarrow 1$, we expect to recover the unweighted HSCM behavior, effectively giving us the same scaling for the $\chi_2$ as for the $\phi_2$ exponent. Thus, we will set $\beta_2$ to $0$ in this limit, and use the above approximation otherwise. This defines our choice for the $\beta_2$ parameter:
\begin{equation}
    \beta_2 \approx
    \begin{cases}
        0,\,\, \eta = 1, \\
        \alpha_2 - 1 + \frac{\eta (\alpha_2 - 1)}{\gamma - 1},\,\, \eta > 1.
    \end{cases}
\end{equation}

\textbf{Finding the $(R, a)$ solution.} As explained in Sec.~\ref{sec:whscm}, we find the $R$, $a$ parameters of the WHSCM by numerically solving Eqs.~(\ref{eq:whscm-average-k}, \ref{eq:whscm-sigma0}) for a fixed $\lambda_0$. For the solver from our code package~\cite{githubcode}, we set $\lambda_0$ such that $\kappa(\lambda_0) = \bar{k}$. This is done to prevent the solver from finding an $(R, a)$ solution that corresponds to a low-degree region ($k < \bar{k}$) of the $\bar{s}(k)$ scaling curve that may not exhibit a clean power-law scaling and distort the resulting baseline for the strength-degree correlation curve. For each solver round, we iteratively update the $\lambda_0$ value corresponding to $\kappa(\lambda_0) = \bar{k}$ with the current solver's guess of $(R, a)$.\\

\section{Behavior of the WHSCM in the $\eta \rightarrow 1$ limit}\label{sec:app-hscm-limit}

With the choice of model parameters above and setting $\eta = 1$, we have the following distribution of the latent parameter $\lambda$:
\begin{equation}\label{eq:hscm-rho-lambda}
    \rho(\lambda) = (\alpha - 1) \lambda^{-\alpha},
\end{equation}
where $\lambda \in (1, \infty)$ and $\alpha = \alpha_1 = \alpha_2 = \gamma$. This gives the expressions for $\kappa(\lambda)$ and $\sigma(\lambda)$:
\begin{equation}
    \kappa(\lambda) = n \int\limits_{1}^{\infty} \frac{(\gamma - 1) \lambda^{\prime -\gamma}}{1 + \frac{2 a e^{2R}}{\lambda \lambda^{\prime}}} d\lambda^{\prime},
\end{equation}
\begin{equation}
    \sigma(\lambda) = n \int\limits_{1}^{\infty} \frac{(\gamma - 1) \lambda^{\prime -\gamma}}{1 + \frac{2 a e^{2R}}{\lambda \lambda^{\prime}}} \cdot \frac{1}{2a} d\lambda^{\prime}.
\end{equation}
The integral for $\kappa(\lambda)$ may be evaluated directly:
\begin{equation}
    \kappa(\lambda) = n \, {_2}F_{1} \left( 1, \gamma - 1, \gamma, -\frac{2 a e^{2R}}{\lambda} \right).
\end{equation}
This function scales linearly with $\lambda$ for large enough $a e^{2R} \gg 1$, therefore, the expected degree $\kappa(\lambda) \sim \lambda$. From the above expressions, we also see that $\sigma(\lambda) = \frac{1}{2a} \kappa(\lambda)$. Thus, the constant $a$ may be easily found given the model input parameter $\sigma_0$, $a = \frac{1}{2 \sigma_0}$. Moreover, the average degree $\bar{k}$ defined by Eq.~\eqref{eq:whscm-average-k} now reads:
\begin{equation}
    \bar{k} = n (\gamma - 1)^2 \Phi\left( -2a e^{2R}, 2, \gamma - 1 \right),
\end{equation}
where $\Phi(z, q_1, q_2)$ is the Lerch transcendent function. These equations allows to solve for the model parameter $R$ numerically, given a target average degree $\bar{k}$. The considered limiting behavior is included as a special case in the code package for generation of WHSCM networks~\cite{githubcode}.

\balance
\onecolumngrid
\twocolumngrid

\end{document}